\documentclass[12pt]{article}

\usepackage{graphicx}
\usepackage{amssymb,latexsym}

\usepackage{xparse}
\usepackage{xspace}
\usepackage{datetime}
\usepackage{amsmath}
\usepackage[labelsep=colon,labelfont=bf]{caption}
\usepackage{tocloft}
\usepackage{pifont}
\usepackage{cite}
\usepackage{color}
\usepackage{collref}
\usepackage[para]{manyfoot}
\usepackage[bottom]{footmisc}
\usepackage{filecontents}
\usepackage[left=2.5cm,right=2.5cm,top=2.5cm,bottom=3cm]{geometry}
\usepackage[linktocpage]{hyperref}

\numberwithin{equation}{section}

\newcommand{\rcite}{\cite}

\NewDocumentCommand\eqn{mo}{%
  \IfNoValueTF{#2}
     {\[ #1 \]}
     {\begin{equation}\label{#2} #1 \end{equation} \expandafter\newcommand\csname #2\endcsname{\eqref{#2}\xspace}\ignorespaces}
}
\NewDocumentCommand\eqna{mo}{%
  \IfNoValueTF{#2}
    {\begin{align*} #1 \end{align*}}
    {\begin{equation}\label{#2}\begin{split} #1 \end{split}\end{equation} \expandafter\def\csname #2\endcsname{\eqref{#2}\xspace}\ignorespaces}
}

\NewDocumentCommand\twoseqn{momoo}{%
    \IfNoValueTF{#5}
       {\begin{subequations}\begin{align} #1\label{#2} \\ #3 \label{#4}  \end{align}\end{subequations} \expandafter\def\csname #2\endcsname{\eqref{#2}\xspace}\ignorespaces \expandafter\def\csname #4\endcsname{\eqref{#4}\xspace}\ignorespaces}
       {\begin{subequations}\label{#5}\begin{align} #1\label{#2} \\ #3 \label{#4}  \end{align}\end{subequations} \expandafter\def\csname #5\endcsname{\eqref{#5}\xspace}\ignorespaces \expandafter\def\csname #2\endcsname{\eqref{#2}\xspace}\ignorespaces \expandafter\def\csname #4\endcsname{\eqref{#4}\xspace}\ignorespaces}
}
\NewDocumentCommand\threeseqn{momomoo}{%
   \IfNoValueTF{#7}
     {\begin{subequations}\begin{align} #1\label{#2} \\ #3 \label{#4} \\ #5 \label{#6} \end{align}\end{subequations} \expandafter\def\csname #2\endcsname{\eqref{#2}\xspace}\ignorespaces \expandafter\def\csname #4\endcsname{\eqref{#4}\xspace}\ignorespaces \expandafter\def\csname #6\endcsname{\eqref{#6}\xspace}\ignorespaces}
     {\begin{subequations}\label{#7}\begin{align} #1\label{#2} \\ #3 \label{#4} \\ #5 \label{#6} \end{align}\end{subequations} \expandafter\def\csname #7\endcsname{\eqref{#7}\xspace}\ignorespaces \expandafter\def\csname #2\endcsname{\eqref{#2}\xspace}\ignorespaces \expandafter\def\csname #4\endcsname{\eqref{#4}\xspace}\ignorespaces \expandafter\def\csname #6\endcsname{\eqref{#6}\xspace}\ignorespaces}
}
\NewDocumentCommand\fourseqn{momomomoo}{%
   \IfNoValueTF{#9}
     {\begin{subequations}\begin{align} #1\label{#2} \\ #3 \label{#4} \\ #5 \label{#6} \\ #7\label{#8} \end{align}\end{subequations} \expandafter\def\csname #2\endcsname{\eqref{#2}\xspace}\ignorespaces \expandafter\def\csname #4\endcsname{\eqref{#4}\xspace}\ignorespaces \expandafter\def\csname #6\endcsname{\eqref{#6}\xspace}\ignorespaces \expandafter\def\csname #8\endcsname{\eqref{#8}\xspace}\ignorespaces}
     {\begin{subequations}\label{#9}\begin{align} #1\label{#2} \\ #3 \label{#4} \\ #5 \label{#6} \\ #7\label{#8} \end{align}\end{subequations} \expandafter\def\csname #9\endcsname{\eqref{#9}\xspace}\ignorespaces \expandafter\def\csname #2\endcsname{\eqref{#2}\xspace}\ignorespaces \expandafter\def\csname #4\endcsname{\eqref{#4}\xspace}\ignorespaces \expandafter\def\csname #6\endcsname{\eqref{#6}\xspace}\ignorespaces \expandafter\def\csname #8\endcsname{\eqref{#8}\xspace}\ignorespaces}
}

\NewDocumentCommand\newsec{mo}{%
  \IfNoValueTF{#2}
     {\section{#1}}
     {\section{#1}\label{#2} \expandafter\gdef\csname #2\endcsname{\ref{#2}\xspace}\ignorespaces}
}
\NewDocumentCommand\subsec{mo}{%
  \IfNoValueTF{#2}
     {\subsection{#1}}
     {\subsection{#1}\label{#2}\expandafter\gdef\csname #2\endcsname{\ref{#2}\xspace}\ignorespaces}
}
\NewDocumentCommand\subsubsec{mo}{%
  \IfNoValueTF{#2}
     {\subsubsection{#1}}
     {\subsubsection{#1}\label{#2}\expandafter\gdef\csname #2\endcsname{\ref{#2}\xspace}\ignorespaces}
}

\makeatletter
\renewcommand\section{\@startsection {section}{1}{\z@}%
{-6ex \@plus -1ex \@minus -.2ex}%
{2.3ex \@plus.2ex}%
{\bfseries}}
\makeatother
\makeatletter
\renewcommand\subsection{\@startsection{subsection}{2}{\z@}%
                                     {-3.25ex\@plus -1ex \@minus -.2ex}%
                                     {1.5ex \@plus .2ex}%
                                     {\itshape}}
\makeatother
\makeatletter
\renewcommand\subsubsection{\@startsection{subsubsection}{3}{\z@}%
                                     {-3.25ex\@plus -1ex \@minus -.2ex}%
                                     {1.5ex \@plus .2ex}%
                                     {\itshape}}
\makeatother
\makeatletter 
\def\@seccntformat#1{\csname the#1\endcsname.\hspace{4.6pt}} 
\makeatother 

\renewcommand{\appendix}{\appendices}

\newenvironment{acknowledgments}{\vspace{12pt}\begin{center}\textbf{Acknowledgments}\end{center}\vspace{-12pt}}{}
\newcommand{\ack}[1]{\begin{samepage}\begin{acknowledgments} {#1} \end{acknowledgments}\end{samepage}}

\collectsep{\ding{70}~}

\newcommand{\foot}{\footnote}
\newdateformat{mydate}{\monthname[\THEMONTH] \THEYEAR} 
\mydate{}

\DeclareNewFootnote[para]{E}[roman]
\newcommand{\email}[1]{\footnoteE{\href{mailto:#1}{\texttt{#1}}}}
\newcommand{\emails}[1]{\let\thefootnote\relax\footnotetext{{\texttt{#1}}}}

\makeatletter
	\renewcommand{\abstract}[1]{\def \@abstract {#1}}
	\newcommand{\affiliation}[1]{\def \@affiliation {#1}}
	\newcommand{\preprint}[1]{\def\@preprint {#1}}
	\abstract{}
	\affiliation{}
	\preprint{}
\makeatother

\makeatletter
\def \maketitle {%
	\begin{titlepage}
	          \begin{flushright}
                           \@preprint
                   \end{flushright}
                           \vspace{2cm}
		\begin{center}
			{\Large\bfseries \@title} 
		
			\bigskip\bigskip\bigskip
		
			\@author 
			
			\bigskip
			
			\emph{\@affiliation}
			
                  	\end{center}
			\bigskip\bigskip

			\noindent\@abstract
			
			\vfill\vfill\vfill\vfill\vfill\vfill\vfill\vfill\vfill\vfill\vfill
			\vfill\vfill\vfill\vfill\vfill\vfill\vfill\vfill\vfill\vfill\vfill

			\noindent\@date
	\end{titlepage}
}
\makeatother

\usepackage{color}

\let\a=\alpha \let\b=\beta \let\g=\gamma \let\d=\delta \let\e=\epsilon
  \let\th=\theta  \let\k=\kappa
\let\l=\lambda \let\m=\mu \let\n=\nu \let\x=\xi \let\p=\pi 
\let\s=\sigma \let\t=\tau    \let\y=\psi
\let\vp=\varphi 
\let\w=\omega      \let\G=\Gamma \let\D=\Delta \let\Th=\Theta \let\L=\Lambda
\let\X=\Xi  \let\S=\Sigma  \let\Y=\Psi
 
\let\la=\label  
  
\def\nn{\nonumber} \def\bd{\begin{document}} \def\ed{\end{document}}
\def\ds{\documentstyle} \let\fr=\frac \let\bl=\bigl \let\br=\bigr
\let\Br=\Bigr \let\Bl=\Bigl
\let\bm=\bibitem
\let\na=\nabla
\def\tU{{\widetilde U}}
\let\pa=\partial \let\ov=\overline
\def\ie{{\it i.e.\ }}
\newcommand{\be}{\begin{equation}}
\newcommand{\ee}{\end{equation}}
\def\ba{\begin{array}}
\def\ea{\end{array}}
\def\ft#1#2{{\textstyle{{\scriptstyle #1}\over {\scriptstyle #2}}}}
\def\fft#1#2{{#1 \over #2}}
\def\F#1#2{{ F_{#1}^{(#2)} }}
\def\cF#1#2{{ {\cal F}_{#1}^{(#2)} }}

\def\R{{\bf R}}
\def\sst#1{{\scriptscriptstyle #1}}
\def\oneone{\rlap 1\mkern4mu{\rm l}}
\def\e7{E_{7(+7)}}
\def\td{\tilde}
\def\wtd{\widetilde}
\def\im{{\rm i}}
\def\bog{Bogomol'nyi\ }
\newcommand{\ho}[1]{$\, ^{#1}$}
\newcommand{\hoch}[1]{$\, ^{#1}$}
\newcommand{\bea}{\begin{eqnarray}}
\newcommand{\eea}{\end{eqnarray}}
\newcommand{\ra}{\rightarrow}
\newcommand{\lra}{\longrightarrow}
\newcommand{\Lra}{\Leftrightarrow}
\newcommand{\ap}{\alpha^\prime}
\newcommand{\bp}{\tilde \beta^\prime}
\newcommand{\cB}{{\cal B}}
\newcommand{\cO}{{\cal O}}
\newcommand{\vecx}{\vec{x}}
\newcommand{\vecy}{\vec{y}}
\newcommand{\vecp}{\vec{p}}
\newcommand{\vecq}{\vec{q}}
\newcommand{\tr}{{\rm tr} }
\newcommand{\Tr}{{\rm Tr} }
\newcommand{\NP}{Nucl. Phys. }

\newcommand{\cL}{{\cal L}}
\newcommand{\cA}{{\cal A}}
\newcommand{\cT}{{\cal T}}
\newcommand{\cD}{{\cal D}}
\newcommand{\cH}{{\cal H}}

\def\sst#1{{\scriptscriptstyle #1}}
\def\0{{\sst{(0)}}}
\def\1{{\sst{(1)}}}
\def\2{{\sst{(2)}}}
\def\3{{\sst{(3)}}}
\def\4{{\sst{(4)}}}
\def\5{{\sst{(5)}}}
\def\6{{\sst{(6)}}}
\def\7{{\sst{(7)}}}
\def\8{{\sst{(8)}}}
\def\9{{\sst{(9)}}}
\def\p{{\sst{(p)}}}
\def\q{{\sst{(q)}}}
\def\ve{\varepsilon}
\def\vf{\varphi}
\def\F{\Phi}
\def\wg{\wedge}

\def\thb{\bar{\theta}}
\def\Thb{\bar{\Theta}}
\def\barp{\bar{p}}
\def\barq{\bar{q}}
\def\barc{\bar{c}}
\def\bard{\bar{d}}
\def\e{\epsilon}

\def \bi{\bibitem}
\def \la {\label}

\def \l {\lambda}
\def\foot{\footnote}
\def \tl  {{\tilde \l}}
\def \sql {{\sqrt \l}}
\def \adss {$AdS_5 \times S^5$\ }
\newcommand{\rf}[1]{(\ref{#1})}
\def \ov {\over}

\def\th{\theta}
\def\Th{\Theta}
\def\vth{\vartheta}
\def\btheta{{\bar\theta}}
\def\ttheta{{{\tilde\theta}}}
\def\bttheta{{{\bar\ttheta}}}
\def\vth{\vartheta}

\def\ra{\rightarrow}
\def\N{\nabla}
\def\F{{\cal F}}
\def\uM{\underline{M}}
\def\uA{\underline{A}}
\def\uN{\underline{N}}
\def\uP{\underline{P}}
\def\ua{\underline{a}}
\def\ub{\underline{b}}
\def\uc{\underline{c}}
\def\ud{\underline{d}}
\def\ue{\underline{e}}
\def\uf{\underline{f}}
\def\ui{\underline{i}}
\def\uj{\underline{j}}
\def\uk{\underline{k}}
\def\ul{\underline{l}}
\def\ual{\underline{\alpha}}
\def\ube{\underline{\beta}}
\def\um{\underline{m}}
\def\un{\underline{n}}
\def\up{\underline{p}}
\def\uq{\underline{q}}
\def\ur{\underline{r}}
\def\us{\underline{s}}
\def\umu{\underline{\mu}}
\def\unu{\underline{\nu}}
\def\ula{\underline{\l}}
\def\uka{\underline{\k}}
\def\usi{\underline{\s}}
\def\urh{\underline{\r}}
\def\cc{\circ}
\def\eqv{\equiv}

\def\ni{\noindent}

\def\Ep{E^{{}^{(+)}}}
\def\Em{E^{{}^{(-)}}}

\def\Mp{M^{{}^{(+)}}}
\def\Mm{M^{{}^{(-)}}}

\def \ha{{1\ov 2}}

\def\r{\rho}

\def\Y{{\rm Y}}
\def\X{{\rm X}}
\def\tY{\tilde{\rm Y}}
\def\tX{\tilde{\rm X}}
\def\dY{\dot{\rm Y}}
\def\dX{\dot{\rm X}}

\def \J {\mathcal{J}}
\def \del {\partial}

\def\dF{\dot{F}}
\def\dG{\dot{G}}
\def\df{\dot{f}}
\def \E {{\cal E}}
\def \S {{\cal S}}
\def \J {{\cal J}}

\def\ms{\mathcal{S}}
\def\mj{\mathcal{J}}
\def\soj{\fr{\ms}{\mj}}
\def \R {{\bf R}}
\def \om {\omega}
\def \bE {\bar E}
\def \x {{\cal X}}

\def \bi{\bibitem}
\def \la {\label}

\def \l {\lambda}
\def\foot{\footnote}
\def \tl  {{\tilde \l}}
\def \sql {{\sqrt \l}}
\def \adss {$AdS_5 \times S^5$\ }
\def \ov {\over}

\def \varpi {{\rm w}}

\def\thb{\bar{\theta}}
\def\Thb{\bar{\Theta}}
\def\mb{\bar{\m}}
\def\ab{\bar{\a}}
\def\zb{\bar{z}}
\def\psib{\bar{\psi}}
\def\barp{\bar{p}}
\def\barq{\bar{q}}
\def\barc{\bar{c}}
\def\bard{\bar{d}}
\def\e{\epsilon}
\def\wb{\bar{w}}
\def\lb{\bar{\l}}
\def\Jb{\bar{J}}
\def\Nb{\bar{N}}
\def\Zb{\bar{Z}}
\def\pab{\bar{\pa}}

\def\At{\tilde{A}}
\def\Bt{\tilde{B}}
\def\Ct{\tilde{C}}
\def\Dt{\tilde{D}}
\def\Et{\tilde{E}}
\def\Ft{\tilde{F}}
\def\Gt{\tilde{G}}
\def\Ht{\tilde{H}}
\def\Mt{\tilde{M}}
\def\Rt{\tilde{R}}
\def\at{\tilde{a}}
\def\bt{\tilde{b}}
\def\ct{\tilde{c}}
\def\dt{\tilde{d}}
\def\et{\tilde{e}}
\def\ft{\tilde{f}}
\def\gt{\tilde{g}}
\def\mt{\tilde{\mu}}
\def\nt{\tilde{\nu}}
\def\vpt{\tilde{\varphi}}

\def\asth{\hat{*}}
\def\phh{\hat{\phi}}

\def\bA{{\bf A}}

\def\ola{\overleftarrow}
\def\ora{\overrightarrow}
\def\alt{\tilde{\a}}

\def\eh{\hat{e}}
\def\eph{\hat{\e}}
\def\ph{\hat{p}}
\def\alh{\hat{\a}}
\def\beh{\hat{\b}}
\def\gah{\hat{\g}}
\def\Fh{\hat{F}}
\def\muh{\hat{\m}}
\def\nuh{\hat{\n}}
\def\thh{\hat{\th}}
\def\dh{\hat{d}}
\def\ih{\hat{i}}
\def\jh{\hat{j}}
\def\kh{\hat{k}}
\def\deh{\hat{\d}}
\def\wh{\hat{w}}
\def\lah{\hat{\l}}
\def\Ah{\hat{A}}
\def\Ch{\hat{C}}
\def\Omh{\hat{\Omega}}

\def\xh{\hat{x}}

\def\ps{\rlap{\, /}\;\,p }
\def\ks{\rlap{\, /}\;\,k }

\def\gym{g_{YM}}

\def\adot{\dot{a}}
\def\bdot{\dot{b}}
\def\bpa{\bar{\pa}}

\def\pr{\prime}
\def\ssk{\medskip}

\title{Tunneling dynamics in exactly-solvable models with triple-well potentials}

\author{V.P. Berezovoj$\,^{\clubsuit,}$\email{berezovoj@kipt.kharkov.ua}, M.I. Konchatnij$\,^{\spadesuit,}$\email{konchatnij@kipt.kharkov.ua} and A.J. Nurmagambetov$\,^{\diamondsuit,}$\email{ajn@kipt.kharkov.ua}}

\affiliation{
A.I. Akhiezer Institute for Theoretical Physics of
NSC KIPT,\\
1 Akademicheskaya St., Kharkov, UA 61108 Ukraine
}

\abstract{
Inspired by new trends in atomtronics, cold atoms devices and Bose-Einstein condensate dynamics, we apply a general technique of N=4 extended Supersymmetric Quantum Mechanics to isospectral Hamiltonians with triple-well potentials, i.e. symmetric and asymmetric. 
Expressions of quantum-mechanical propagators, which take into account all states of the spectrum, are obtained, within the N = 4 SQM approach, in the closed form. For the initial Hamiltonian of a harmonic oscillator, we obtain the explicit expressions of potentials, wavefunctions and propagators. The obtained results are applied to tunneling dynamics of localized states in triple-well potentials and for studying its features. In particular, we observe a Josephson-type tunneling transition of a wave packet, the effect of its partial trapping and a non-monotonic dependence of tunneling dynamics on the shape of a three-well potential. We investigate, among others, the possibility of controlling tunneling transport by changing parameters of the central well, and we briefly discuss potential applications of this aspect to atomtronic devices.

\phantom{We study, among others, possibility to manage tunneling transport by changing parameters of central well, and briefly discuss potential applications of this aspect to atomtronic devices.}
{PACS numbers: 12.60.Jv, 03.65.-w, 03.65.Xp 
}

}

\begin{document}
\maketitle


\newsec{Introduction}

Studies of tunneling dynamics in quantum models with triple-well potentials are becoming attractive in view of new trends in atomtronics associated with, as intensively discussed in the literature 
\rcite{Seaman07},\rcite{Muga04},\rcite{Pepino09},\rcite{Micheli04},\rcite{Stickney07},\rcite{Benseny10},\rcite{Muga07},\rcite{Muga06},\rcite{Gajdacz12}, atomic diodes and transistors. The prospects of creating new types of devices are opened up with developing experimental tools for manipulations with cold atoms and for laser monitoring and control 
\rcite{Rab07},\rcite{Viscondi11},\rcite{Rab12},\rcite{Tan12} 
of  tunneling transport. A triple-well structure has been used in experiments of light transfer in a triple-well optical waveguide \rcite{Longhi06},\rcite{Longhi07}, classically modelling more subtle quantum coherent transport of neutral atoms or electrons in multi-well traps.

Theoretical investments in atomtronic devices are also made in the studies of tunneling transport in one-dimensional (1D) quantum-mechanical models, or models of Gross-Pitaevsky type, with double- or triple-well potentials. In these studies, properties of tunneling dynamics, such as the particle transition between side wells with negligible filling of the central well, one way transport, a feedback of a small central well filling on the tunneling rate, etc \rcite{Creentree04},\rcite{Eckert04},\rcite{Opartny09},\rcite{Lu10},\rcite{Lu11},\rcite{Gerritsma11}, are the focus. The resonance tunneling effect in the triple-well 1D time-independent models and their experimental observations are also discussed in \rcite{Deunff10},\rcite{Deunff12}. However, the analysis is carried out on exotic triple-well potentials, which are constructed by combining separate (step-like or parabolic-type) wells or by separating wells with delta-function-type barriers. As a consequence, spectra and wavefunctions in such potentials are recovered by numerical, i.e. non-analytical, methods. This makes it difficult to study effects in the tunneling dynamics associated with changes in the potential shape (e.g., with deformation of one well out of a number of wells)
\footnote{See, however, \rcite{Sato02}, \rcite{Sato05}, \rcite{Cao11} (and Refs. therein) for other types of approches. }.

The investigation of the space-time evolution of quantum mechanical systems with potentials with several local minima is one of the most complicated and important tasks. The complexity of the problem forces one to use various simplifications/approximations, e.g., matrix Hamiltonians with account of few lowest levels \rcite{Kuklinski89},\rcite{Marte91},\rcite{Bergmann98},\rcite{Chen12}, or purely numerical simulations in phenomenological models. These simplifications can be used for figuring out the common properties of tunneling dynamics, but not for the studies of subtle effects. However, the detailed consideration of tunneling dynamics is becoming important in condensed matter physics, where the possibility of constructing new atomtronic devices based on cold atoms is being intensively explored. Therefore, studying the subtleties of tunneling dynamics in multi-well potentials requires using models with controllable characteristics, while the studies of space-time dynamics in tunneling processes requires going beyond the standard approximations (that take into account just a few of the lowest states of the Hamiltonian and deal with parameters chosen according to the potential shape). The above-mentioned requirements are inherent to the exactly solvable models of extended supersymmetric quantum mechanics (SQM) with multi-well potentials \rcite{Andrianov93},\rcite{Andrianov95},\rcite{Plyuschay00},\rcite{Andrianov12},\rcite{Berezovoj10},\rcite{Berezovoj12}.

Extended N = 4 SQM allows one to construct an exactly solvable model with a multi- well potential, properties of which -- eigenvalues of lowest states of the spectrum, degree of deformation of the potential, etc -- can be preassigned. The technique developed in \rcite{Jauslin88},\rcite{Pupasov07} can be used to derive the explicit analytical expression of the corresponding propagator. Hence, it is quite natural to extend the methods proposed in \rcite{Berezovoj10},\rcite{Berezovoj12} to models with triple-well potentials. Free parameters of exactly solvable models within N = 4 extended SQM make it possible to vary the shape of the potentials, and hence the corresponding propagators, in a wide range and, in a consistent manner, to study the effects of these changes on the dynamics of tunneling transitions. Recall that exact propagators take into account {\it all states} of N = 4 SQM Hamiltonians, which allows one to study the tunneling dynamics of wave packages beyond the approximations of phenomenological approaches.

The remainder of this paper is organized as follows. In section 2, we outline the procedure for constructing exactly solvable isospectral Hamiltonians of N = 4 SQM with three-well potentials and discuss their general properties. Previously obtained results in \rcite{Berezovoj12} are crucial to this end. In what follows, we focus on the harmonic oscillator model and use the harmonic oscillator Hamiltonian as a starting point in constructing a family of three-well potentials with different degrees of deformation. The corresponding exact propagators are obtained in section 3. The obtained results are applied in section 4 to study features of the tunneling dynamics of the originally localized states in symmetric and asymmetric potentials. Systems with three-well potentials have a rich quantum dynamics, and we, in particular, observe a Josephson- type tunneling transition of a wave packet, the effect of its partial trapping and a non-monotonic dependence of tunneling dynamics on the shape of a three-well potential. We collect our conclusions in section 5, where we also discuss the prospects for further development and application of our current results.

\newsec{Constructing isospectral Hamiltonians with triple-well potentials}

\subsec{Warm up exercise: a double-well potential}

The procedure of constructing Hamiltonians with exactly-solvable N-well potentials from the initial Hamiltonian $H_0$, the spectrum and the wave functions of which are known, is based on adding complementary levels below the $H_0$ ground-state level with the energy $E_0$. This procedure has been applied to the harmonic oscillator Hamiltonian in \cite{Zheng84} to construct new isospectral two-well Hamiltonians and was further developed in \cite{Jauslin88}. Recall that in constructing models with multi-well potentials within the approach proposed in \cite{Zheng84,Jauslin88}, the original Hamiltonian $H_0=p^2/2+U(x)$ (in $\hbar=m=1$ units) implies the asymptotic behavior of the potential $U(x\ra\pm \infty)\sim C_{\pm}|x|^{\eta_{\pm}}$ \cite{Jauslin88} with the non-negative constants $C_\pm, \eta_\pm$. This class of Hamiltonians involves models with purely discrete spectra (for positive $C_\pm, \eta_\pm$) and also with mixed, discrete and continuous, spectra (once one or more constants $C_\pm, \eta_\pm$ are equal to zero). The mathematical basis of this procedure, which is known as the Crum-Krein method, was proposed in the original papers \rcite{Crum55},\rcite{Krein57}, and intensively discussed in subsequent publications (see, e.g., \cite{Jauslin88,Pupasov07,Odake11}, and a pedagogical introduction into the Darboux transformations in quantum mechanics and SQM \cite{Williams08}). The Crum-Krein method is equivalent to constructing isospectral Hamiltonians in the polynomial SQM (reducible case) \cite{Andrianov93,Andrianov95,Plyuschay00,Andrianov12}, which is characterized by the successive application of supersymmetry transformations to the initial Hamiltonian.

To construct Hamiltonians with triple-well potentials, we adopt the previously used procedure \rcite{Berezovoj10,Berezovoj12} of obtaining exactly solvable models of extended SQM with double-well potentials. Note that the formalism of extended SQM is not the only way to achive this; the same results can also be obtained by the use of the Crum-Krein method. However, some steps of the Crum-Krein procedure, e.g., the normalization of extra states wavefunctions, require a separate treatment with additional care. That is why we prefer to follow the method of successive application of supersymmetry transformations to the initial Hamiltonian $H_0$, which leads directly to the isospectral Hamiltonians with multi-well potentials.

Let us briefly recall the main results of \rcite{Berezovoj10,Berezovoj12}, which will be used in what follows. Suppose, we have an exactly-solvable (non-supersymmetric) model with Hamiltonian $H_0=p^2/2+U(x)$ (in $\hbar=m=1$ units). For definiteness, let us fix $H_0$ to be a Hamiltonian with a purely discrete spectrum. In N = 4 SQM, adding an extra level with the energy $\ve$ below the ground-state energy $E_0$ of the initial Hamiltonian $H_0$, i.e. $\ve < E_0$, one may construct a new isospectral Hamiltonian $H_-^-$
\foot{Recall, that there are four Hamiltonians, $H_+^+$, $H^+_-$, $H^-_+$ and $H^-_-$, in N=4 extended SQM, three of which, i.e. $H_+^+$, $H^-_+$ and $H^-_-$, are related to each other by SUSY transformations (see, e.g., \cite{Berezovoj91} for details). } and its wave functions:
\eqna{
&H^-_-=\left(H_0-\ve \right)-\fr{d^2}{dx^2}\ln \varphi(x,\ve,c),\qquad H_0-\ve\equiv H^-_+ ,\\
&\psi^-_-(x,E_i)=\fr1{\sqrt{2(E_i-\ve)}}\fr{{\text{W}}\Big(\psi^-_+ (x,E_i),\varphi(x,\ve,c)\Big)}{\varphi(x,\ve,c)}\,,\\
&\psi^-_-(x,E=0)=\fr{N^{-1}}{\varphi(x,\ve,c)}\equiv \vpt(x,\ve,c).
}[H--2w]
Here, $\y^-_+(x,E_i)$ are the normalized wavefunctions that correspond to the states of the initial Hamiltonian $H_0$ with energies $E_i$ (energy is counted off $\ve$); ${\text{W}}\Big(f_1(x),f_2(x) \Big)=f_1 f^\pr_2-f^\pr_1 f_2$ is the Wronskian of two functions $f_{1,2}$; $f^\pr \equiv df/dx$. $\vp$ and $\vpt$ are new functions that appear in the following way. The solution to the equation $H_0\, \vp(x,\ve)=\ve \vp(x,\ve)$ for $\ve< E_0$ consists of two	linearly independent non-negative functions $\vp_I(x,\ve), I=1,2$ with the following asymptotic behavior: $\vp_1(x\ra -\infty)\ra +\infty, \vp_2(x\ra -\infty)\ra 0$ and $\vp_1(x\ra +\infty)\ra 0, \vp_2(x\ra +\infty)\ra +\infty$. The existence of corresponding states for potentials considered here was proved in \cite{Jauslin88}, where the analysis of their properties was also made. Solutions to the Schr\"odinger equation $H_0 \vf_I(x)=\ve \vf_I(x)$ do not  belong to the spectrum, since $\vf_I(x) \notin L^2$. Following \cite{Jauslin88}, it is always possible to choose the integration constant $c$ in such a way that the general solution $\vp(x,\ve,c)=N\left(\vp_1(x,\ve)+c\vp_2(x,\ve)\right)$ does not have any nodes on the whole axis.  Taking the latter into account, the function $\vpt (x,\ve,c)=N^{-1}/\vp(x,\ve,c)$ is finite and can be normalized to unity for every value of parameters $\ve$ and $c$. In other words, the spectrum of $H^-_-$ contains the additional, with respect to the $H_0$ spectrum, state with the energy $\ve$ and the wave function $\tilde{\vf}(x,\ve,c)\in L^2$. The normalization of the wavefunctions of excited states of $H^-_-$ is preserved. For a specific choice of parameters $\ve$ and $c$, the function $\vpt (x,\ve,c)$, and hence the potential of $H^-_-$, has local maxima and minima. The double-well structure of the potentials $U^-_-(x,\ve,c)$ becomes sharp under $(E_0-\ve)/E_0 \ll 1$.

The super-Hamiltonian of N=4 SQM $H^{\s_1}_{\s_2} (\s_{1,2}=\pm)$ \rcite{Berezovoj10},\rcite{Berezovoj12} also contains the $H^+_+$ part, that gets the extra level in the spectrum. Explicitly, the Hamiltonian $H^+_+$ and its wave functions are
\eqna{
H^+_+=(H_0-\ve)-&\fr{d^2}{dx^2}\ln \left(\fr{\vpt(x,\ve,c)}{1+\l \int^x_{x_i} d\xi \,\vpt^2 (\xi,\ve,c)}\right),\quad \y^+_+(x,E=0)=\fr{N^{-1}_\l \,\vpt (x,\ve,c)}{(1+\l \int^x_{x_i} d\xi \,\vpt^2 (\xi,\ve,c))},\\
&\y^+_+ (x,E_i)=\fr1{\sqrt{2(E_i-\ve)}}
\left(\fr{d}{dx}-\fr{d}{dx}\ln \fr{\vpt(x,\ve,c)}{(1+\l \int^x_{x_i} d\xi \,\vpt^2 (\xi,\ve,c))}\right)\y^-_+(x,E_i).
}[H++2w]
Here, $\l$ is the integrations constant, characterizing the general solution to the Riccati equation (see \rcite{Berezovoj91} for details), whose value is restricted to $\l > -1$. Note that the potentials of $H^-_-$ and $H^+_+$ are shape invariant \rcite{Gendenshtein83}, i.e. $U^-_-$ and $U^+_+$ are related to each other by non-coordinate transformations of their parameters.

Fixing $c=1$ in \rf{H--2w} and \rf{H++2w} and using the relation \rcite{Jeffreys66}\footnote{In \rf{Rel} $A_1$, $A_2$ are the arbitrary coefficients (non-negative in our case), $x$ and $x_i$ are the integration limits ($x_i$ may be chosen to be equal to $-\infty$ for the class of potentials considered here), and $y_{1,2}(x)$ and $y_{1,2}(x_i)$ are the values of functions $y_{1,2}$ at $x$ and $x_i$.}
\eqn{
\int^x_{x_i} d\xi \, \fr{{\text{W}}\Big(y_1(\xi),y_2(\xi)\Big)}{\left( A_1 y_1(\xi)+A_2 y_2(\xi)\right)^2}=-\fr1{\left(A_1^2+A_2^2 \right)}\left[\left(\fr{A_2 y_1(x)-A_1 y_2(x)}{A_1 y_1(x)+A_2 y_2(x)}\right) -\left(\fr{A_2 y_1(x_i)-A_1 y_2(x_i)}{A_1 y_1(x_i)+A_2 y_2(x_i)}\right)\right]
}[Rel]
we obtain, after inserting \rf{Rel} in \rf{H--2w} and \rf{H++2w},
\eqna{
&H^-_-=H^-_+ -\fr{d^2}{dx^2}\ln \left(\vp_1(x,\ve)+\vp_2(x,\ve)\right),\quad H^+_+=H^-_+ - \fr{d^2}{dx^2}\ln \left(\vp_1(x,\ve)+\L(\ve,\l)\vp_2(x,\ve) \right),\\
&\L(\ve,\l)=\fr{\Delta(\infty,\ve,c=1)-\l-(\l+1)\Delta(-\infty,\ve,c=1)}{\Delta(\infty,\ve,c=1)+\l-(\l+1)\Delta(-\infty,\ve,c=1)},\quad \Delta(x,\ve,c)=\fr{c\,\vp_1(x,\ve)-\vp_2(x,\ve)}{\vp_1(x,\ve)+c\,\vp_2(x,\ve)},\\
&\psi^-_-(x,E_i)=\fr1{\sqrt{2(E_i-\ve)}}\fr{{\text{W}}\Big(\psi^-_+ (x,E_i),\varphi(x,\ve,1)\Big)}{\varphi(x,\ve,1)}\,,\quad
\psi^-_-(x,E=0)=\fr{N^{-1}}{\varphi(x,\ve,1)}\equiv \vpt(x,\ve,1)\,, \\
&\psi^+_+(x,E_i)=\fr1{\sqrt{2(E_i-\ve)}}\fr{{\text{W}}\Big( \psi^-_+(x,E_i),\vp(x,\ve,\L)\Big)}{\vp(x,\ve,\L)},\quad \psi^+_+(x,E=0)=\fr{N^{-1}_\L}{\vp(x,\ve,\L)}\equiv \vpt (x,\ve,\L) .
}[H--++2w]
Clearly, the potential of $H^-_-$ is defined by the symmetric combination $\vp_1(x,\ve)+\vp_2(x,\ve)$ (recall, $\vp_I$, $I=1,2$ are the linearly-independent solutions to the equation $H_0\vp_I(x,\ve)=\ve \vp_I(x,\ve)$ with $\ve<E_0$), whilst $U^+_+$ is defined by the asymmetric combination $\vp_1(x,\ve)+\L(\ve,\l)\vp_2(x,\ve)$. The same concerns the wave functions of excited states of $H^-_-$ and $H^+_+$.

Equation \rf{Rel} is used to compute the normalization constants $N^{-1}$ and $N^{-1}_\L$ of the corresponding ground-state wave functions. They are related to each other by $N^{-2}_\L=\L N^{-2}=(\l+1)N^{-2}$, where
\eqn{
N^{-2}=-\fr{2{\text{W}} \Big(\vp_1,\vp_2 \Big)}{\Delta(+\infty,\ve,c=1)-\Delta(-\infty,\ve,c=1)}={\text{W}}\Big(\vp_1(x,\ve),\vp_2 (x,\ve)\Big).
}[N-22w]

For the initial Hamiltonian $H_0$ of the harmonic oscillator model, we obtain $\vp_1(\xi,\bar{\ve})=D_\n (\sqrt{2}\xi)$, $\vp_2(\xi,\bar{\ve})=D_\n (-\sqrt{2}\xi)$, ($\xi=\sqrt{\w}x, \n=-1/2 + \bar{\ve}, \bar{\ve}=\ve/\w$; $D_\n(x)$ is the parabolic cylinder function; $D_\n (\sqrt{2}\xi)$ and $D_\n (-\sqrt{2}\xi)$ are linearly independent of the non-integer $\n$ that is supposed in what follows), and ${\text{W}}\Big(\vp_1,\vp_2\Big)=2\sqrt{\pi \w}/\G(-\n)$ \rcite{Gradshtein94}
with the Gamma-function $\G(-\n)$. The normalization constant $N$ is
\[
N^{-2}=-\fr{{\text{W}}\Big(\vp_1,\vp_2\Big)}{\Delta(+\infty,\bar{\ve},1)}=\fr{2\sqrt{\pi \w}}{\G(-\n)}\, .
\]
Note that the only way of varying the shape of the potential in terms of dimensionless variables $\xi$ is to vary parameters $\L$ and $\bar{\ve}$ in the following range $0 < \bar{\ve} < 1/2$, $\L > 0$. In natural units $x$, there is one more parameter $\w$, the variation of which may change the potential shape, and, in particular, the positions of its local minima.

\subsec{Exactly-solvable Hamiltonians with triple-well potentials}

Now, we are ready to proceed with constructing isospectral Hamiltonians with triple-well potentials. Equations \rf{H--2w}, \rf{H++2w}, \rf{H--++2w} are the basic equations to this end. 

Let us consider a double-well Hamiltonian $H^+_+$ as the initial one (cf \rf{H--++2w}):
\eqn{
\Ht_0-\ve_1\equiv H^+_+=(H_0-\ve)-\fr{d^2}{dx^2}\ln \big(\vp_1(x,\ve)+\L(\ve,\l)\vp_2(x,\ve)\big).
}[Ht03w] 
The spectrum of $\Ht_0$ apparently contains the whole spectrum of the original Hamiltonian $H_0$, and the extra level with the energy $\ve <E_0$, which becomes the ground state of the Hamiltonian $\Ht_0$. Clearly, $\Ht_0$ contains a deformed potential $\tilde{U}^+_+(x,\ve,\L(\ve,\l))$.

Next, consider solutions to the equation $\Ht_0 \,\chi(x)=\ve_1 \chi(x)$ with $\ve_1 <\ve$. As in the previous case of double-well Hamiltonians, one finds two linearly-independent (non-normalizable) solutions to this equation:
\eqn{
\chi_1(x,\ve_1,\ve,\L)=\fr{{\text{W}}\Big(\vp_1(x,\ve_1),\vp(x,\ve,\L)\Big)}{\vp(x,\ve,\L)},\quad \chi_2(x,\ve_1,\ve,\L)=-\fr{{\text{W}}\Big(\vp_2(x,\ve_1),\vp(x,\ve,\L)\Big)}{\vp(x,\ve,\L)}\,,
}[chi123w]
where $\vp_I(x,\ve_1), I=1,2$ are the solutions to equations $H_0\, \vp_I(x,\ve_1)=\ve_1 \vp_I(x,\ve_1)$ with the original Hamiltonian $H_0$, and $\vp(x,\ve,\L)=\vp_1(x,\ve)+\L(\ve,\l)\vp_2(x,\ve)$.

Solutions \rf{chi123w} are non-negative, and have the following asymptotes: $\chi_1(x\ra -\infty) \ra +\infty, \chi_2(x\ra -\infty)\ra 0$ and $\chi_1(x\ra +\infty)\ra 0, \chi_2(x \ra +\infty)\ra +\infty$. The minus sign in the expression of $\chi_2(x,\ve_1,\ve,\L)$ provides the correct asymptotic behavior. For the specific choice of integration constants, the general solution
to $\Ht_0 \,\chi(x,\ve_1,\ve,\L)=\ve_1 \chi(x,\ve_1,\ve,\L)$, $\ve_1 <\ve$,
\[
\chi(x,\ve_1,\ve,\L,\tilde{c})=\tilde{N}(\chi_1(x,\ve_1,\ve,\L)+\ct \chi_2 (x,\ve_1,\ve,\L))
\]
does not have any nodes on the whole axis. 

In what follows, we set the constant $\ct$ to unity. Then, the function
\[
\tilde{\chi}(x,\ve_1,\ve,\L,\ct=1)=\fr{\tilde{N}^{-1}}{\chi_1(x,\ve_1,\ve,\L)+\chi_2(x,\ve_1,\ve,\L)}
\]
is finite, and it can be normalized to unity for every specific choice of parameters.

To construct a three-well Hamiltonian from the two-well Hamiltonian $\Ht_0$, consider
\eqna{
\Ht_0-\ve_1\equiv H^+_+ =\Ht^-_+(x,p)=\fr12 p^2+\fr12 \Big[(V(x,\ve_1,\ve,\L))^2+V^{\pr}(x,\ve_1,\ve,\L)\Big],\\
V(x,\ve_1,\ve,\L)=\fr{d}{dx}\ln \chi(x,\ve_1,\ve,\L,1)=\fr{d}{dx}\ln \fr{\text{W} \Big(\phi (x,\ve_1,1),\vp(x,\ve,\L) \Big)}{\vp(x,\e,\L)} \, .
}[Ht0V3w]
Here 
\[
 V=\fr{d}{dx}\left(\tilde{W}-\fr12 \ln \tilde{W}^\pr \right), \quad
\tilde{W}(x)=-\fr12 \ln \left(1+\tilde{\l} \int_{x_i} d\xi \, \tilde{\chi}^2(\xi,\ve_1,\ve,1) \right) 
\]
with the superpotential $\tilde{W}$ (see \rcite{Berezovoj91} for details), and $\phi(x,\ve_1,1)=\vp_1(x,\ve_1)-\vp_2(x,\ve_1)$.

The spectrum of $\Ht^-_+$ completely coincides with that of $H^+_+$, and has the extra level with $\ve < E_0$ in comparison to the spectrum of $H_0$. Recall, that the Hamiltonian $H^+_+$ \rf{H--++2w} (or \rf{Ht0V3w}) has the potential with a double-well shape. 

The superpartner of $\Ht^-_+$ is a triple-well Hamiltonian
\eqn{
\Ht^-_-=\Ht^-_+ -\fr{d^2}{dx^2}\ln \chi(x,\ve_1,\ve,\L,1)=(H_0-\ve_1)-\fr{d^2}{dx^2}\ln {\text{W}}\Big(\phi (x,\ve_1,1),\vp(x,\ve,\L)\Big) .
}[Ht--3w]
Its spectrum consists of the states of $H_0$, and two extra levels with the energies $\ve$ and $\ve_1$  ($E_0>\ve > \ve_1$). The corresponding wavefunctions to these extra levels are
\threeseqn{
\Psi_0 (x,\ve_1;\ve,\L)=&\fr{\tilde{N}^{-1}}{\chi_1(x,\ve_1,\ve,\L)+\chi_2(x,\ve_1,\ve,\L)}=\fr{\tilde{N}^{-1}\vp(x,\ve,\L)}{{\text{W}}\Big(\phi(x,\ve_1,1),\vp(x,\ve,\L)\Big)}\, ,
}[]
{
\Psi_1(x,\ve;\ve_1,\L)=&\fr{N^{-1}_{\L}}{\sqrt{2(\ve-\ve_1)}}\left(\fr{d}{dx}-\fr{\chi^\pr(x,\ve_1,\ve,\L,1)}{\chi(x,\ve_1,\ve,\L,1)} \right) \vpt (x,\ve,\L)=\fr{\sqrt{2(\ve-\ve_1)}N^{-1}_{\L} \phi(x,\ve_1,1)}{{\text{W}}\Big(\phi(x,\ve_1,1),\vp(x,\ve,\L)\Big)} \, ,
}[Psi1]{&\Psi^-_-(x,E_i)=\fr1{\sqrt{2(E_i-\ve_1)}}\left(\fr{d}{dx}-\fr{\chi^\pr(x,\ve_1,\ve,\L,1)}{\chi(x,\ve_1,\ve,\L,1)} \right) \psi^+_+(x,E_i) \nn\\ 
&=\sqrt{\fr{E_i-\ve}{E_i-\ve_1}}\left(\psi^-_+(x,E_i)+\fr{\ve-\ve_1}{E_i-\ve_1} \phi(x,\ve_1,1)\fr{{\text{W}}\Big(\psi^-_+(x,E_i),\vp(x,\ve,\L)\Big) }
{{\text{W}}\Big(\phi(x,\ve_1,1),\vp(x,\ve,\L)\Big)}\right) \,.
}[Psi2][wf2extra]
Here, $\chi(x,\ve_1,\ve,\L,1)\equiv \chi(x,\ve_1,\ve,\L,\tilde{c}=1)=\tilde{N}(\chi_1(x,\ve_1,\ve,\L)+\chi_2 (x,\ve_1,\ve,\L))$.

By the use of \rf{Rel} and the asymptotic behavior of $\chi_I(x,\ve_1,\ve,\L), I=1,2$, it is easy to calculate the normalization constant of the ground state wave function $\Psi_0(x,\ve_1,\ve,\L)$:
\eqn{
\tilde{N}^{-2}={\text{W}}\Big(\chi_1,\chi_2\Big)\equiv 2(\ve-\ve_1) {\text{W}}\Big(\vp_1(x,\ve_1),\vp_2(x,\ve_1)\Big)\,.
}[Ntilde3w]
To obtain a more symmetric representation of the extra levels wave functions, let us introduce another normalizations constant, $N^{-2}_{\L}\ra 2(\ve-\ve_1)N^{-2}_{\L}$. Then, equation \rf{Psi1} becomes
\[
\Psi_1(x,\ve,\ve_1,\L)=\fr{N^{-1}_\L \phi(x,\ve_1,1)} {{\text{W}}\Big(\phi(x,\ve_1,1),\vp(x,\ve,\L)\Big)} \,.
\]

Recall that the solutions $\vp_I(x,\ve_1), I=1,2$, satisfy \rf{Ht0V3w}, \rf{Ht--3w}, \rf{wf2extra} through $\phi(x,\ve_1,1)=\vp_1(x,\ve_1)-\vp_2(x,\ve_1)\equiv \phi(x,\ve_1,\L_1=1)$. Below we show that fixing the starting Hamiltonian $H_0$ to the harmonic oscillator one, a symmetric potential $\tilde{U}^-_-$ of $\Ht^-_-$ corresponds to the choice $\L=\L_1=1$. Deformations of $\tilde{U}^-_-$, even if $\L_1=1$, mainly affect side wells and to lesser extent the central well.

\subsection{Shape-invariance of triple-well potentials in N=4 SQM isospectral Hamiltonians}

Let us establish the shape-invariance of the potentials satisfying $\Ht^-_-$ and $\Ht^+_+$ (see \rcite{Berezovoj10},\rcite{Berezovoj12} for a double-well case). $\Ht^-_-$ and $\Ht^-_+ =\Ht_0-\ve_1$ are related via (cf. \rf{H++2w})
\eqn{
\Ht^+_+ = (\Ht_0-\ve_1)-\fr{d^2}{dx^2}\ln \left(\fr{\tilde{\chi}(x,\ve_1,\ve,\L,1)}{1+\tilde{\l}\int^x_{x_i} d\xi \, \tilde{\chi}^2 (\xi,\ve_1,\ve,\L,1)}\right) \,.
}[Ht++toHt--]
Using \rf{Rel}, one can obtain
\[
\fr{\tilde{\chi}(x,\ve_1,\ve,\L,1)}{1+\tilde{\l}\int^x_{x_i} d\xi \, \tilde{\chi}^2 (\xi,\ve_1,\ve,\L,1)}=\fr{\tilde{N}^{-1}_{\L_1}}{\chi_1(x,\ve_1,\ve,\L)+(\tilde{\l}+1)\chi_2(x,\ve_1,\ve,\L)},~~\tilde{N}^{-2}_{\L_1}=(\tilde{\l}+1)\tilde{N}^{-2}\,.
\]
As a result,
\eqn{
\Ht^+_+=\Ht^-_+ -\fr{d^2}{dx^2}\ln |\chi(x,\ve_1,\ve,\L_1,\L)|=(H_0-\ve_1)-\fr{d^2}{dx^2}\ln {\text{W}}\Big(\phi(x,\ve_1,\L_1),\vp(x,\ve,\L)\Big)\,,
}[Ht++toH0]
where $\L_1=\tilde{\l}+1$ ($\tilde{\l}> -1$ for normalized wave functions), and  $\chi(x,\ve_1,\ve,\L_1,\L)=\tilde{N}(\chi_1(x,\ve_1,\ve,\L)+\L_1 \chi_2(x,\ve_1,\ve,\L))$. Comparing eq. \rf{Ht++toH0} with eq. \rf{Ht--3w}, one can conclude the shape invariance of $\Ht^-_-$ and $\Ht^+_+$. 

The wavefunctions of $\Ht^+_+$ states come from the wavefunctions of $\Ht^-_-$ states \rf{Psi2}, replacing $\phi(x,\ve_1,1)$ with $\phi(x,\ve_1,\L_1)=\vp_1(x,\ve_1)-\L_1\vp_2(x,\ve_1)$:

\threeseqn{
&\tilde{\Psi}_0 (x,\ve_1,\L_1;\ve,\L)=\fr{\tilde{N}^{-1}_{\L_1}\vp(x,\ve,\L)}{{\text{W}}\Big(\phi(x,\ve_1,\L_1),\vp(x,\ve,\L)\Big)}\, ,
}[Psi03w]
{
&\tilde{\Psi}_1(x,\ve,\L;\ve_1,\L_1)=\fr{N^{-1}_{\L} \phi(x,\ve_1,\L_1)}{{\text{W}}\Big(\phi(x,\ve_1,\L_1),\vp(x,\ve,\L)\Big)} \, ,
}[Psi13w]{\Psi^+_+(x,E_i)=&\sqrt{\fr{E_i-\ve}{E_i-\ve_1}}\left(\psi^-_+(x,E_i)+\fr{\ve-\ve_1}{E_i-\ve_1} \phi(x,\ve_1,\L_1)\fr{{\text{W}}\Big(\psi^-_+(x,E_i),\vp(x,\ve,\L)\Big) }
{{\text{W}}\Big(\phi(x,\ve_1,\L_1),\vp(x,\ve,\L)\Big)}\right) \,.
}[][wtHt++3w]

Two remarks are now in order. Equations \rf{Ht--3w}, \rf{wf2extra} and \rf{Ht++toH0}, \rf{wtHt++3w} include the wave functions of $H_0$, as well as non-normalizable solutions to the Schr\"odinger equation $H_0 \vp(x)=\ve \vp(x)$ with $\ve <E_0$. 
And potentials satisfying $\Ht^-_-$ and $\Ht^+_+$ depend on a number of parameters $(\ve_1,\ve,\L_1,\L)$ that makes it possible to vary their shapes, from purely symmetric to essentially deformed.

\subsection{Triple-well model from the harmonic oscillator Hamiltonian}

Consider the initial Hamiltonian to be the harmonic oscillator one: 
\[
H_0=\fr{p^2}{2}+\fr{\w^2 x^2}{2}\, .
\]
As it has been noted above, the non-normalized solutions of $H_0 \vp(x)=\ve \vp(x), \ve<E_0$ are the parabolic cylinder functions
\[
\vp_1(\xi,\bar{\ve})=D_\n (\sqrt{2}\xi),\quad \vp_2(\xi,\bar{\ve})=D_\n (-\sqrt{2}\xi),\qquad \xi=\sqrt{\w}x, ~\n=-\fr12+\bar{\ve}, ~\bar{\ve}=\fr{\ve}{\w}\,,
\]
the Wronskian of which is
\[
{\text{W}}\Big(\vp_1,\vp_2\Big)=\fr{2\sqrt{\pi \w}}{\G(-\n)} \,.
\]
Then, the normalization constants are
\[
N^{-2}=4(\n-\m)\fr{\sqrt{\pi\w}}{\G(-\n)},\quad N^{-2}_\L=\L N^{-2},\qquad \L>0,~\n=-\fr12+\bar{\ve} \,,
\]
and
\[
\tilde{N}^{-2}=4(\n-\m)\fr{\sqrt{\pi\w}}{\G(-\m)},\quad N^{-2}_{\L_1}=\L_1 \tilde{N}^{-2},\qquad \L_1>0,~\m=-\fr12+\bar{\ve}_1,~ \bar{\ve}_1=\fr{\ve_1}{\w}\,.
\]

Figure \rf{fig:image1} shows the shape of the potential $\tilde{U}^-_-(\xi,\m,\n,\L)$ for the change in the energy of the ground state for a fully symmetric case (see the left plot in figure \rf{fig:image1}), and for a non-trivial value of the deformation parameter of the side wells (see the right plot in figure \rf{fig:image1}). For a symmetric potential, increasing the value of $\ve_1$ enlarges the width of the central local minimum and reduces the widths of the side local minima. The distance between the side wells also gets increased. Hence, the energy $\ve_1$ is the parameter which controls the shape of potential. Note, that varying $\ve_1$ does not affect the energies of the first and the second excited states. Having a non-trivial deformation ($\L<1$), decreasing $\ve_1$ results in the growth of the barrier in front of the deeper minimum.

\begin{figure}[h]
\begin{minipage}[h]{0.499\linewidth}
\center{\includegraphics[width=0.99\linewidth]{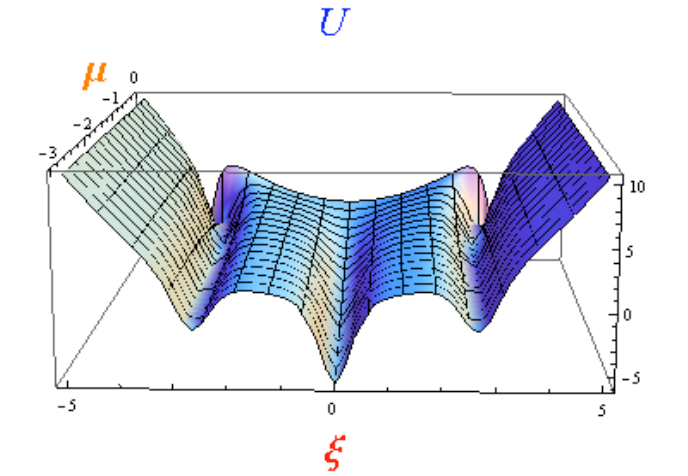} \\ a)}
\end{minipage}
\hfill
\begin{minipage}[h]{0.499\linewidth}
\center{\includegraphics[width=1.2\linewidth]{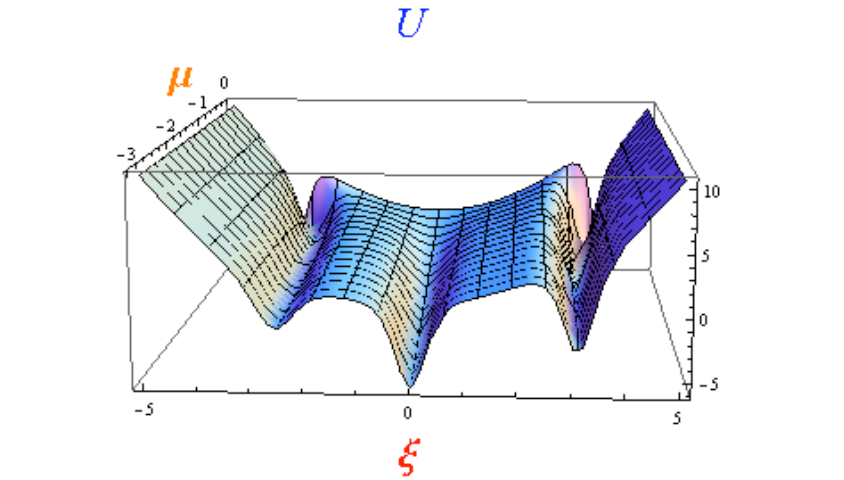} \\ b)}
\end{minipage}
\caption{$\tilde{U}^-_-(\xi,\m,\n,\L)$ shape variation for: a) $\L=1,\n=-0.02$; b) $\L=0.05,\n=-0.02$.   }
\label{fig:image1}
\end{figure}

It is important to trace back the behavior of wave functions corresponding to the lowest states of $\Ht^-_-$ (under-barrier states). When all wells have the same depth (see the left panel in figure \rf{fig:image2}), one may note that the ground-state wavefunction (solid curve) is almost entirely concentrated in the central well and is completely absent in the side wells. While the wavefunctions of the first excited state (dash-dotted curve) and the second excited state (dotted curve) are almost zero in the central well. Their values in the under-barrier region is enough to provide the tunneling process. In the case of $\L \ne 1$, the wave functions distribution does not essentially change in the central well.   A redistribution of the wavefunctions of the first and the second excited states turns out in the side wells: the right well essentially contains the first excited level wavefunction, while the second excited level wavefunction is mostly concentrated in the left well (see the right panel in figure \rf{fig:image2}).

\begin{figure}[h]
\begin{minipage}[h]{0.49\linewidth}
\center{\includegraphics[width=0.9\linewidth]{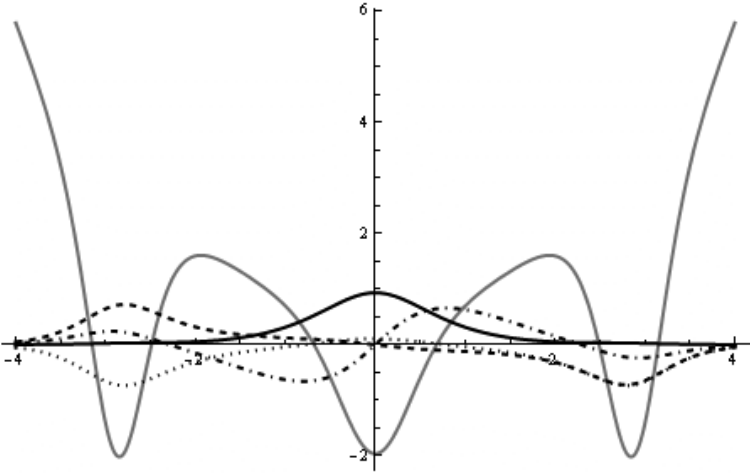} \\ a)}
\end{minipage}
\hfill
\begin{minipage}[h]{0.49\linewidth}
\center{\includegraphics[width=0.99\linewidth]{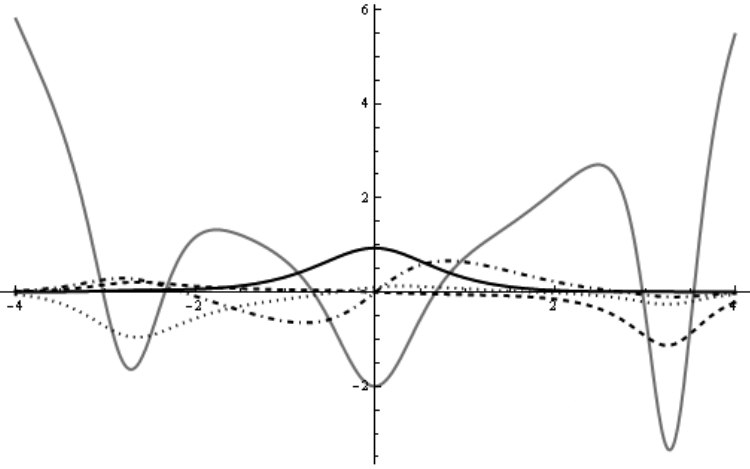} \\ b)}
\end{minipage}
\caption{$\tilde{U}^-_-(\xi,\m,\n,\L)$ and wavefunctions for: a) $\L=1$, $\m=-1$, $\n=-0.02$; b) $\L=0.05$, $\m=-1$, $\n=-0.02$.   }
\label{fig:image2}
\end{figure}

A more detailed analysis of the behavior of $\Ht^-_-$ and $\Ht^+_+$ wavefunctions is postponed to Section 4.

\newsec{Quantum-mechanical propagator in systems with three-well potentials}

The study of temporal evolution in quantum-mechanical systems with multi-well potentials is one of the most important tasks of quantum mechanics. Due to the complexity of tunneling dynamics in real systems, one has to use different simplifications or approximations, such as the two-mode approximation, or phenomenological matrix models based on account of a few levels. These models can be used to study the general properties of tunneling dynamics, leaving aside more subtle quantum-mechanical effects.

A more general way to describe the space-time evolution in quantum-mechanical systems is to use the quantum-mechanical propagator. Recall that the exact propagator contains contributions of all states of the quantum-mechanical Hamiltonian. Suppose that a quantum-mechanical object, say, a wave packet with the squeeze parameter $R$,
\eqn{
\Phi(x_0,0)=\left(\fr{\w e^{2R}}{\pi}\right)^{1/4}\exp \left(-\fr{\w}2 (x_0-a_0)^2 e^{2R} \right) \,,
}[Phiinit]
is initially, i.e. at $t=0$, localized at $a_0$. Then, its space-time evolution is described by \rcite{Feynman65}
\eqn{
\Phi(x,t)=\int^{+\infty}_{-\infty} dx_0\, K(x,t;x_0,0) \Phi(x_0,0),\quad K(x,t;x_0,0)=\sum^\infty_{n=0} \psi_n(x) \psi^{*}_n(x_0)e^{-iE_n t} \,.
}[evol]
Here $K(x,t;x_0,0)$ is the propagator, the explicit form of which allows one to study the dynamics of localized states in potentials of any complexity. Currently, there are only a few examples of exactly-solvable models with explicitly derived expressions of propagators $K(x,t;x_0,0)$ \rcite{Grosche98}. As a rule, they correspond to Hamiltonians with one-well potentials.

A procedure of getting new models with propagators related to those of exactly-solvable models with one-well potentials was proposed in \rcite{Jauslin88},\rcite{Pupasov07}. Following  \rcite{Jauslin88},\rcite{Pupasov07}, the exact expressions of propagators for quantum-mechanical systems with two-wells potentials were derived by two of us in \rcite{Berezovoj12}, and the results
applied to study the tunneling dynamics of localized states. Here we extend the method of \rcite{Berezovoj12} to Hamiltonians with triple-well potentials.

Let us introduce the propagators $\tilde{K}^-_-$ and $\tilde{K}^+_+$ corresponding to the Hamiltonians $\Ht^-_-$ \rf{Ht--3w} and $\Ht^+_+$ \rf{Ht++toH0}.
The explicit form of $\tilde{K}^+_+$ can be obtained on account of the shape-invariance established above. Having the expression of $\tilde{K}^+_+$,  $\tilde{K}^-_-$ is derived by setting $\L_1=1$ in the expression of $\tilde{K}^+_+$. The latter is
\eqna{
&\tilde{K}^+_+(x,t;x_0,0)=\sum_n \Psi^+_+(x,E_n)\Psi^+_+(x_0,E_n)e^{-iE_n t} \\
+\tilde{\Psi}_1(x,\ve,\L;\ve_1,\L_1)\tilde{\Psi}_1(x_0,\ve,\L;\ve_1,\L_1)&e^{-i\ve t}
+\tilde{\Psi}_0(x,\ve_1,\L_1;\ve,\L)\tilde{\Psi}_0(x_0,\ve_1,\L_1;\ve,\L)e^{-i\ve_1 t} 
\, ,
}[K++3w]
where the summation (in the absence of the continuous spectrum) is carried over the states, related to the initial Hamiltonian $H_0$. Other terms of \rf{K++3w} correspond to two extra levels with energies $\ve$ and $\ve_1$ ($\ve > \ve_1)$. To establish the relation of the propagator \rf{K++3w} to that of the exactly-solvable model with the Hamiltonian $H_0$ we will consider only the first term on the l.h.s. of \rf{K++3w}. The wavefunctions of $\Ht^+_+$ and $H_0$ are related to each other with
\eqn{
\Psi^+_+(x,E_n)=\fr1{2\sqrt{(E_n-\ve)(E_n-\ve_1)}}\,\tilde{L}_x L_x \,\psi^-_+(x,E_n), \quad E_n \ne \ve,\ve_1\,,
}[PsiLL]
where $L_x$ and $\tilde{L}_x$ are the N=4 SQM intertwining operators
\[
L_x=\fr{d}{dx}-\fr{d}{dx}\ln \vp(x,\ve,\L),
\]
\[
\tilde{L}_x=\fr{d}{dx}+\fr{d}{dx} \ln \tilde{\Psi}_0 (x,\ve_1,\L_1;\ve,\L)=\fr{d}{dx}+\fr{d}{dx} \ln \fr{\vp(x,\ve,\L)}{{\text{W}}\Big(\phi(x,\ve_1,\L_1),\vp(x,\ve,\L)\Big)} \,.
\]
By use of \rf{PsiLL} the first term on the l.h.s. of \rf{K++3w} can be written as
\eqn{
\sum_n \Psi^+_+(x,E_n)\Psi^+_+(x_0,E_n)e^{-iE_n t}=\tilde{L}_x \tilde{L}_{x_0}L_x L_{x_0}\sum_n \fr1{4 (E_n-\ve)(E_n-\e_1)}\, \psi^-_+(x,E_n)\psi^-_+(x_0,E_n)e^{-iE_n t}
}[Psi++L4]

On account of the identity
\[
\fr1{(E_n-\ve)(E_n-\ve_1)}=\fr1{(\ve-\ve_1)}\left(\fr1{E_n-\ve}-\fr1{E_n-\ve_1}\right)\,,
\]
and a property of a propagator $K_0 (x,t;z,0)$
\[
\psi^-_+(x,E_n)e^{-iE_n t}=\int^{+\infty}_{-\infty} dz \, K_0 (x,t;z,0) \, \psi^-_+(z,E_n) \,,
\]
one obtains
\eqn{
\sum_n \Psi^+_+(x,E_n)\Psi^+_+(x_0,E_n)e^{-iE_n t}=\fr{\tilde{L}_x \tilde{L}_{x_0}L_x L_{x_0}}{4(\ve-\ve_1)} \int^{+\infty}_{-\infty} dz \, K_0 (x,t;z,0) \left(G_0(z,x_0;\ve)-G_0(z,x_0;\ve_1) \right)\,,
}[Psi++L4G0]
where
\[
G_0(z,x_0;\ve)=\sum_n \fr{\psi^-_+(z,E_n)\psi^-_+(x_0,E_n)}{E_n-\ve}
\]
is the Green function of $H_0$. It is well known (see, e.g., \rcite{Morse53}), the Green function admits the following representation in terms of solutions to the Schr\"odinger equation in the interval $[a,b]$:
\eqn{
G^-_+(x,y,\ve)=-\fr2{{\text{W}}\Big(f_l,f_r\Big)}\left(f_l(x,\ve)f_r(y,\ve)\th (y-x)+(f_l(y,\ve)f_r(x,\ve)\th (x-y)\right)\, ,
}[G-+]
satisfying the boundary conditions $f_l (a,\ve)=0$ and $f_r (b,\ve)=0$.
In our conventions $f_l(x,\ve)=\vp_2(x,\ve)$, $f_r(x,\ve)=\vp_1(x,\ve)$; the same is for $\ve_1$.

Substituting \rf{G-+} into \rf{Psi++L4G0} and acting with $\tilde{L}_{x_0}L_{x_0}$, after some tedious calculations, we obtain
\eqna{
\sum_n \Psi^+_+(x,E_n)\Psi^+_+(x_0,E_n)e^{-iE_nt}=\tilde{L}_x L_x \left(F_1(x,x_0,\ve,\L,\ve_1,\L_1)+F_2(x,x_0,\ve,\L,\ve_1,\L_1) \right),\\
F_1(x,x_0,\ve,\L,\ve_1,\L_1)=-\fr1{2(\ve-\ve_1)}K_0(x,t;x_0,0)-\fr{\phi(x_0,\ve_1\L_1)}{{\text{W}} \Big(\phi(x_0,\ve_1,\L_1),\vp(x_0,\ve,\L)\Big)} \times \\
\times \int^{+\infty}_{-\infty} dz 
\, K_0(x,t;z,0)\left(\L \phi_2(z,\ve)\th (x_0-z)-\vp_1(z,\ve)\th(z-x_0) \right)\,,\\
F_2(x,x_0,\ve,\L,\ve_1,\L_1)=\fr1{2(\ve-\ve_1)}K_0(x,t;x_0,0)-\fr{\varphi(x_0,\ve_1\L_1)}{{\text{W}} \Big(\phi(x_0,\ve_1,\L_1),\vp(x_0,\ve,\L)\Big)} \times \\
\times \int^{+\infty}_{-\infty} dz 
\, K_0(x,t;z,0)\left(\L_1 \varphi_2(z,\ve_1)\th (x_0-z)+\vp_1(z,\ve_1)\th(z-x_0) \right)\,.
}[F1F2]
Note, that every term on the l.h.s. of \rf{F1F2} 
contains $K_0(x,t;x_0,0)$, but with the opposite signs. As a result, these terms compensate each other, and we avoid the need to operate (differentiate) with a singular function (distribution). The multiplier in front of the integral in $F_{1}$ ($F_2$) coincides, modulo the normalization constant, with the wavefunction of the first excited (of the ground state) level of $\Ht^+_+$. Therefore, the final expression of the propagator has the following form:
\eqna{
&\tilde{K}^+_+(x,t;x_0,0)=-\fr{\tilde{\Psi}_1(x_0,\ve,\L;\ve_1,\L_1)}{N^{-1}_{\L}}\tilde{L}_x L_x \times \\
&\times \int^{+\infty}_{-\infty} dz \, K_0(x,t;z,0)\left(\L \vp_2(z,\ve)\th (x_0-z)-\vp_1(z,\ve)\th (z-x_0) \right) \\
&-\fr{\tilde{\Psi}_0(x_0,\ve_1,\L_1;\ve,\L)}{\tilde{N}^{-1}_{\L_1}}\tilde{L}_x L_x \times \\
&\times \int^{+\infty}_{-\infty} dz \, K_0(x,t;z,0)\left(\L_1 \vp_2(z,\ve_1)\th (x_0-z)+\vp_1(z,\ve_1)\th (z-x_0) \right)\\
&+\tilde{\Psi}_1(x,\ve,\L;\ve_1,\L_1)\tilde{\Psi}_1(x_0,\ve,\L;\ve_1,\L_1)e^{-i\ve t}+\tilde{\Psi}_0(x,\ve_1,\L_1;\ve,\L)\tilde{\Psi}_0(x_0,\ve_1,\L_1;\ve,\L)e^{-i\ve_1 t} \,.
}[Kt++fin]

Choosing the initial exactly-solvable Hamiltonian $H_0$ we have to specify the solutions $\vp_I(x,\ve),I=1,2$, and the propagator $K_0(x,t;x_0,0)$. For the harmonic oscillator model $\vp_1(\xi,\bar{\ve})=D_\n(\sqrt{2}\xi)$, $\vp_2(\xi,\bar{\ve})=D_\n(-\sqrt{2}\xi)$, and $K_0(x,t;x_0,0)$ is as follows \rcite{Grosche98}:
\eqn{
K_0(x,t;x_0,0)=\left( \fr{\w e^{-i\pi(\fr12+n)}}{2\pi \sin \w\t}\right)^{1/2} \exp \left(\fr{i\w}{2\sin \w t} \left[(x^2+x_0^2)\cos \w t - 2xx_0 \right] \right)\,,
}[K0HO]
where $t=n\pi/\w+\t, n\in \mathbb{N}$, $0<\t<\pi/\w$. The presence of $\exp (-i\pi(\fr12+n))$ in \rf{K0HO} provides the correct behavior of the propagator for all values of $t$.

\newsec{Dynamics of localized states in multi-well potentials}

Now, we are ready to study features of the localized states dynamics in three-well potentials, with taking into account the contribution of all states forming a localized state $\Phi(x,0)$. The basic equations are equations \rf{wtHt++3w}, \rf{evol}, \rf{Kt++fin}. Changing the parameters $(\ve,\ve_1,\L,\L_1)$ enables us to investigate the effect of the potential shape on the tunneling dynamics in a wide range. To specify the consideration, we will fix the initial Hamiltonian to the Hamiltonian of the harmonic oscillator model in terms of dimensionless variable $\xi=\sqrt{\w} x$. Clearly, changing the value of $\w$ results in the appropriate change of the potential shape (in particular, in the localization of the side minima) in natural units $x$. Furthermore, it is interesting to study the features of the tunneling dynamics with dependence on the choice of $\Phi(x,0)$. This especially becomes true for the amplification of tunneling transitions by the use of a small part of the initial packet in the central well.

In general, dynamics of localized states cannot be correctly described with taking into account just a few lowest under-barrier states. The contribution of higher excited states in forming of $\Phi(x,0)$ becomes essential during the tunneling process, and their role increases with the degree of localization. The approach that we follow enables us to consider the features of the dynamics of localized states by taking into account all states of an exactly solvable Hamiltonian with symmetric and asymmetric multi-well potentials.

To proceed further, let us write down the basic equations \rf{wtHt++3w}, \rf{evol}, \rf{Kt++fin} in terms of dimensionless units $\xi$. We have
\eqn{
\Phi(\xi,T)=\int^{+\infty}_{-\infty} d\zeta \, \tilde{K}^+_+ (\xi,T;\zeta,0)\Phi(\zeta,0)\,,
}[Phidiml]
where $\Phi(\xi,0)$ is the wave packet at the initial time, and
\eqna{
&\tilde{K}^+_+(\xi,T;\xi_0,0)=-\fr{\tilde{\Psi}_1(\xi_0,\ve,\L;\ve_1,\L_1)}{N^{-1}_{\L}}\tilde{L}_{\xi} L_{\xi} \times \\
&\times \int^{+\infty}_{-\infty} d\zeta \, K_0(\xi,T;\zeta,0)\left(\L D_\n (-\sqrt{2}\zeta)\th (\xi_0-\zeta)-D_\n (\sqrt{2}\zeta)\th (\zeta-\xi_0) \right) \\
&-\fr{\tilde{\Psi}_0(\xi_0,\ve_1,\L_1;\ve,\L)}{\tilde{N}^{-1}_{\L_1}}\tilde{L}_{\xi} L_{\xi} \times \\
&\times \int^{+\infty}_{-\infty} d\zeta \, K_0(\xi,T;\zeta,0)\left(\L_1 D_\m (-\sqrt{2}\zeta)\th (\xi_0-\zeta)+D_\m (\sqrt{2}\zeta)\th (\zeta-\xi_0) \right)\\
&+\tilde{\Psi}_1(\xi,\ve,\L;\ve_1,\L_1)\tilde{\Psi}_1(\xi_0,\ve,\L;\ve_1,\L_1)e^{-i\ve T}+\tilde{\Psi}_0(\xi,\ve_1,\L_1;\ve,\L)\tilde{\Psi}_0(\xi_0,\ve_1,\L_1;\ve,\L)e^{-i\ve_1 T} 
}[Kt++HO]
with
\eqn{
K_0(\xi,T;\zeta,0)=\left( \fr{\w e^{-i\pi(\fr12+n)}}{2\pi \sin \t}\right)^{1/2} \exp \left(\fr{i}{2\sin T} \left[(\xi^2+\zeta^2)\cos T - 2\xi\zeta \right] \right)\,,
}[K0THO]
$T=\pi n+\t, n\in \mathbb{N}, 0<\t<\pi$. The wavefunctions of the ground and the first excited states \rf{Psi03w}, \rf{Psi13w} are, in terms of dimensionless units,
\twoseqn{
\tilde{\Psi}_0 (\xi,\ve_1,\L_1;\ve,\L)=\fr{\tilde{N}^{-1}_{\L_1}\vp(\xi,\ve,\L)}{{\text{W}}\Big(\phi(\xi,\ve_1,\L_1),\vp(\xi,\ve,\L)\Big)}\, ,}[]
{
\tilde{\Psi}_1(\xi,\ve,\L;\ve_1,\L_1)=\fr{N^{-1}_{\L} \phi(\xi,\ve_1,\L_1)}{{\text{W}}\Big(\phi(\xi,\ve_1,\L_1),\vp(\xi,\ve,\L)\Big)} \, ,
}[][Psit01xi]
where $\tilde{N}^{-2}_{\L_1}=\L_1 \tilde{N}^{-2}$, $\tilde{N}^{-2}_{\L}=\L \tilde{N}^{-2}$,
\[
\tilde{N}^{-2}=4(\n-\m)\fr{\sqrt{\pi\w}}{\G(-\m)},\qquad {N}^{-2}=4(\n-\m)\fr{\sqrt{\pi\w}}{\G(-\n)} \,,
\]
$\L,\L_1 >0$, $\m=-1/2 + \ve_1/\w$, $\n=-1/2 + \ve/\w$. The functions $\phi$ and $\vp$, satisfying \rf{Psit01xi}, are as follows
\[
\phi(\xi,\ve_1,\L_1)=D_\m (\sqrt{2}\xi)-\L_1 D_\m(-\sqrt{2}\xi),\qquad 
\varphi(\xi,\ve,\L)=D_\n (\sqrt{2}\xi)+\L D_\n(-\sqrt{2}\xi)\,.
\]
Now equations \rf{Phidiml}, \rf{Kt++HO}, \rf{K0THO} and \rf{Psit01xi} become the basic equations to study the features of the tunneling dynamics in a wide range of varied parameters ($\ve,\ve_1,\L,\L_1,\w$). 

We started with the wave packet initially localized in one of side wells
\[
\Phi(\zeta,0)=\left(\fr{e^{2R}}{\pi} \right)^{1/4} \exp \big( -\fr12 (\zeta-\zeta_i)^2 e^{2R} \big) \,,
\]
where $\zeta_i$ is the position of the corresponding well minimum. Next, we figure out the effect of the deformation degree on the tunneling dynamics, and its dependence on the energies
of extra levels in $\tilde{U}^+_+(\xi,\ve,\L,\ve_1,\L_1)$. Looking at figure \rf{fig:image1}, one may note that the three-well
structure is most evident for the choice of parameters $\ve,\ve_1$, when the state with the energy $\ve$ forms the tunnel doublet with the initial Hamiltonian $H_0$ and the ground-state energy $E_0$, $\D =(E_0-\ve)/\w \ll 1$, and the energy $\ve_1$ is fixed to be $|\ve_1| \simeq \w$. (In what follows, we consider $\w=1$. This frequency value will be silently supposed in calculations of numerics in tables 1 and 2.) Therefore, it is advisable to fix the value of $\D$ and to study effects of changing $\ve_1$ on the tunneling dynamics. Decreasing the ground state energy of $\Ht^{\s_2}_{\s_1}$ results in deepening of the central well and in concentration of the ground-state wavefunction therein. However, the ground-state wavefunction of  $\Ht^{\s_2}_{\s_1}$ still plays an essential role in the tunneling dynamics, since transition rates from side wells depend on the overlap integrals between the ground-state wavefunction and the wavefunctions of other states. In this respect, changes in $\ve_1$ with fixed values of $\D$ enable us to change the tunneling rate. Concerning deformations of the potential, we will mainly consider them by the use of $\L$ variations. It leads mostly to changes in the shape of the side wells, though a deformation of the central well  also takes place (see figure \rf{fig:image2}).

\subsec{Symmetric case}

All states of Hamiltonians $\Ht^-_-$ and $\Ht^+_+$ form the complete and orthonormal basis. We would like to show the efficiency of the basis in the series expansion of $\Phi(\xi,0)$ over the lowest states in symmetric and asymmetric potentials (states counting starts from the ground state), comparing the exact results of \rf{Phidiml} with the following approximation 
\eqn{
|\Phi(\xi,T)|=\left | \sum^{n_{max}}_{n=0} \, c_n \tilde{\psi}^{(-)+}_{(-)+} (\xi,E_n) \exp (-iE_n T) \right | .
}[Phiapp]
The value of squeeze parameter $R$ (cf. \rf{Phiinit})  is fixed by the requirement of the minimal number of states giving the essential contribution to the series expansion of $\Phi(\xi,0)$.

Table \rf{tab1} contains the coefficients of the $\Phi(\xi,0)$ series expansion over the states of $\Ht(\xi,\m,\n,\L=1)$ for different choices of $(\m,\n,R)$ (recall that here $\w=1$). `Left' indicates the minimum in which the wave packet is initially localized.  

\begin{table}[h] \hspace{-1cm}
			\begin{tabular}{|c|c|c|c|c|c|c|c|c|c|}
		\hline
			 state number & 1  & 2 & 3 & 4 & 5 & 6 & 7 & 8 & 9 \\ \hline
			 $\L=1,R=1.0$, Left & $0.56$ & $0.699$ & $- 0.417$ & $0.038$ & $-0.048$ & $0.06$ & $-0.067$ & $0.062$ & $-0.042$ \\ 
			  $\m=-0.03,\n=-0.02$ &  &  &  &  &  &  &  &  &  \\ \hline
			$\L=1,R=1.0$, Left & $0.54$ & $0.698$ & $- 0.441$ & $0.041$ & $-0.052$ & $0.066$ & $-0.072$ & $0.066$ & $-0.045$ \\ 
			$\m=-0.032,\n=-0.02$ &  &  &  &  &  &  &  &  &  \\ \hline
			$\L=1,R=0.66$, Left & $0.038$ & $0.68$ & $- 0.69$ & $0.162$ & $-0.138$ & $0.098$ & $-0.051$ & $0.016$ & $-0.003$ \\ 
			$\m=-1,\n=-0.02$ &  &  &  &  &  &  &  &  &  \\ \hline
			$\L=1,R=0.6$, Left & $0.016$ & $0.68$ & $- 0.69$ & $0.168$ & $-0.134$ & $0.085$ & $-0.039$ & $0.015$ & $-0.012$ \\  
			$\m=-2.0,\n=-0.02$ &  &  &  &  &  &  &  &  &  \\ \hline
		\end{tabular}
		\caption{\label{tab1} Coefficients of the $\Phi(\xi,0)$ series expansion \rf{Phiapp} over the states of $\Ht(\xi,\m,\n,\L)$.}
		\label{tab1}
	\end{table}
It is important to note that once the energies of extra states are close to the ground-state energy of the harmonic oscillator model, the main contribution to the series expansion of $\Phi(\xi,0)$ comes from the three lowest states. This result is in favor of the application of phenomenological three-level Hamiltonians to tunneling processes at fixed parameters of the potential \rcite{Kuklinski89,Marte91,Bergmann98,Chen12}. Considered in Table \rf{tab1}, the values of parameters correspond to a wide central well (see the left plot in figure \rf{fig:image1}) and to the strong overlapping of wavefunctions, which leads to the appearance of a sufficient portion of the wave packet in the central well in the tunneling process (see figure \rf{fig:image3}). 
\begin{figure}[h]
\begin{minipage}[h]{0.499\linewidth}
\center{\includegraphics[width=1.\linewidth]{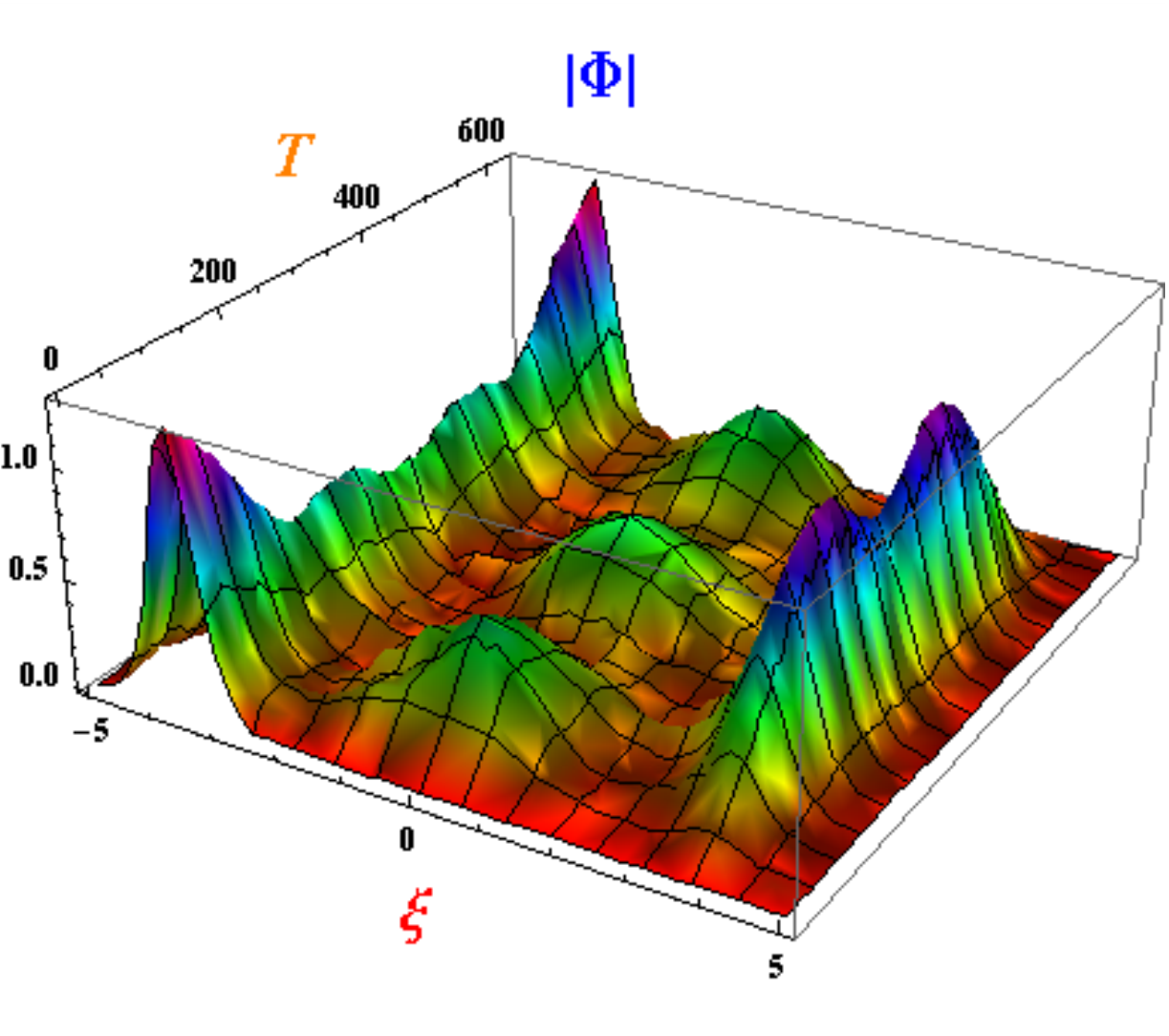} \\ a)}
\end{minipage}
\hfill
\begin{minipage}[h]{0.49\linewidth}
\center{\includegraphics[width=1.1\linewidth]{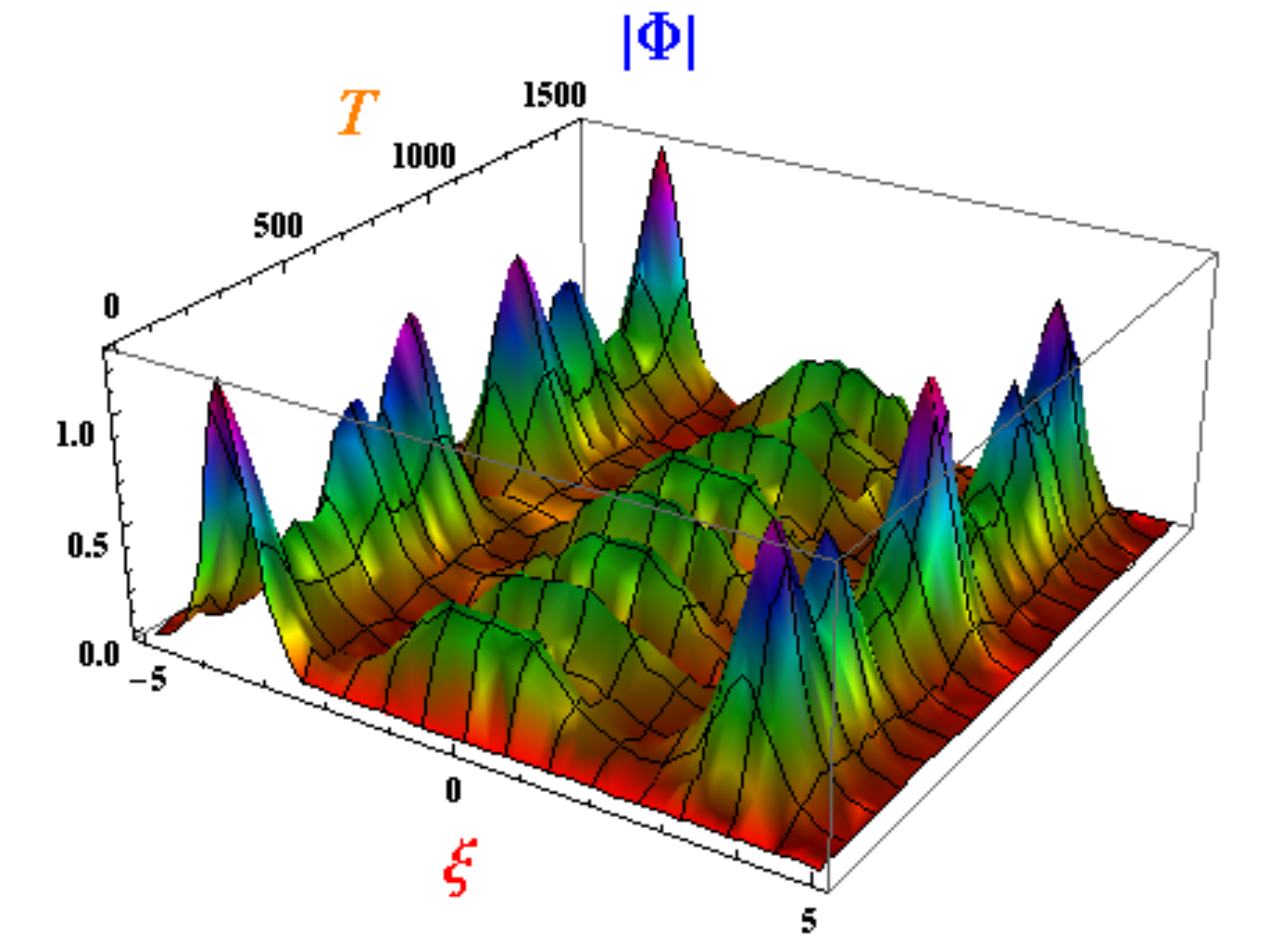} \\ b)}
\end{minipage}
\caption{$|\Phi(\xi,T)|$ dependence in the symmetric case: plot a) corresponds to the first row of table \rf{tab1} and plot b) corresponds to the second row of table \rf{tab1}.   }
\label{fig:image3}
\end{figure}

It is worth mentioning that a small change in the $\m$ value (the second row of table \rf{tab1}) results in an essential rearrangement of the dynamics of tunnel transitions. This is indicated by increasing the revival time of the wave packet, as well as by the more complicated structure of $|\Phi(\xi,T)|$. Therefore, we can conclude the non-monotonic dependence of the tunneling dynamics under the changing properties of a three-well potential. It means that the values of parameters of phenomenological models \rcite{Kuklinski89,Marte91,Bergmann98,Chen12} should also have a non-monotonic behavior. The explicit behavior of probability density in the central well and its sharp dependence on a small change of the well shape parameters can be found in figure \rf{fig:image4}.
\begin{figure}[h]
\begin{minipage}[h]{0.49\linewidth}
\center{\includegraphics[width=0.9\linewidth]{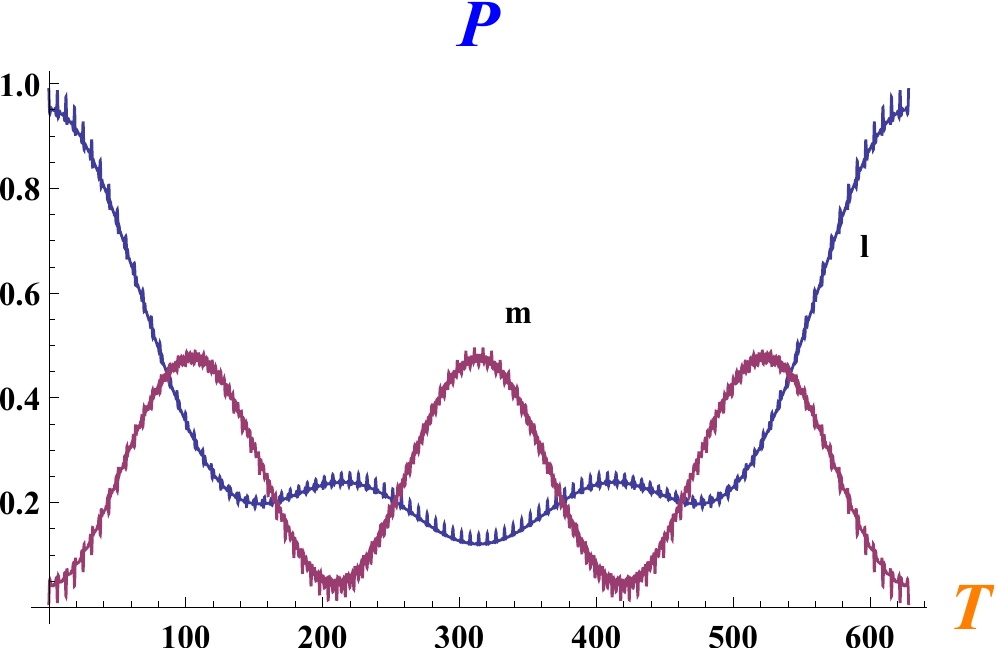} \\ a)}
\end{minipage}
\hfill
\begin{minipage}[h]{0.499\linewidth}
\center{\includegraphics[width=0.9\linewidth]{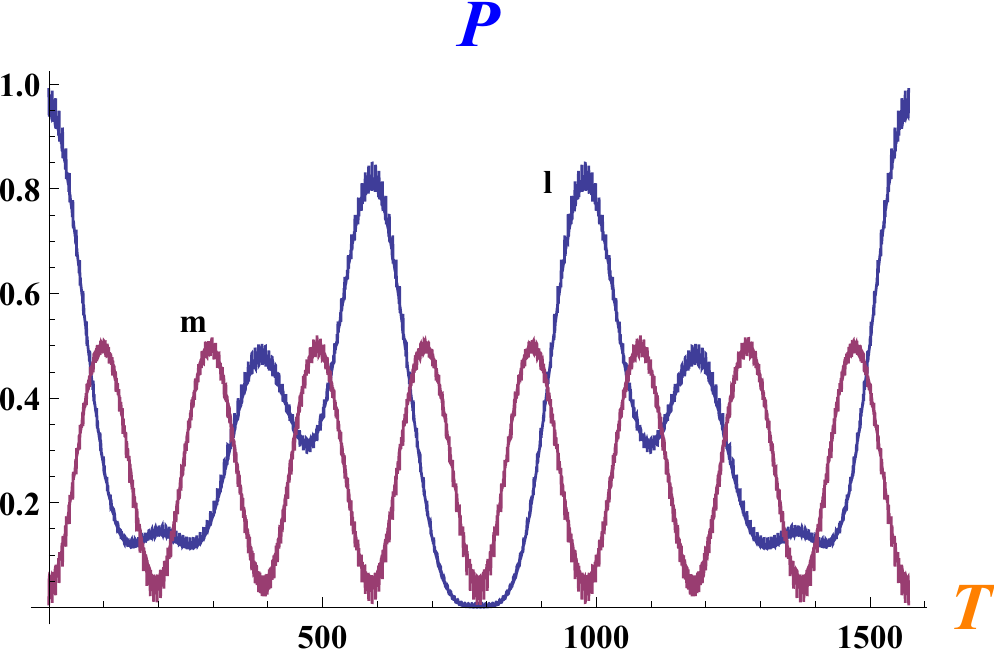} \\ b)}
\end{minipage}
\caption{Probability densities in the central (m-curve) and in the left (l-curve) wells: plot a) corresponds to $\n=-0.02$, $\m=-0.03$, and plot b) corresponds to $\n=-0.02$, $\m=-0.032$.   }
\label{fig:image4}
\end{figure}

When the ground-state energy $|\ve_1| \simeq \w$, the contribution of the ground-state wavefunction in the series expansion of $\Phi(\xi,0)$ 
is very small (cf. third and fourth rows if table \rf{tab1}). Indeed, looking at figure \rf{fig:image2}, one may note that the ground-state wavefunction is located at the central well, and its value in the side wells is negligibly small (see figure \rf{fig:image5}). 
\begin{figure}[h]
\begin{minipage}[h]{0.49\linewidth}
\center{\includegraphics[width=1.0\linewidth]{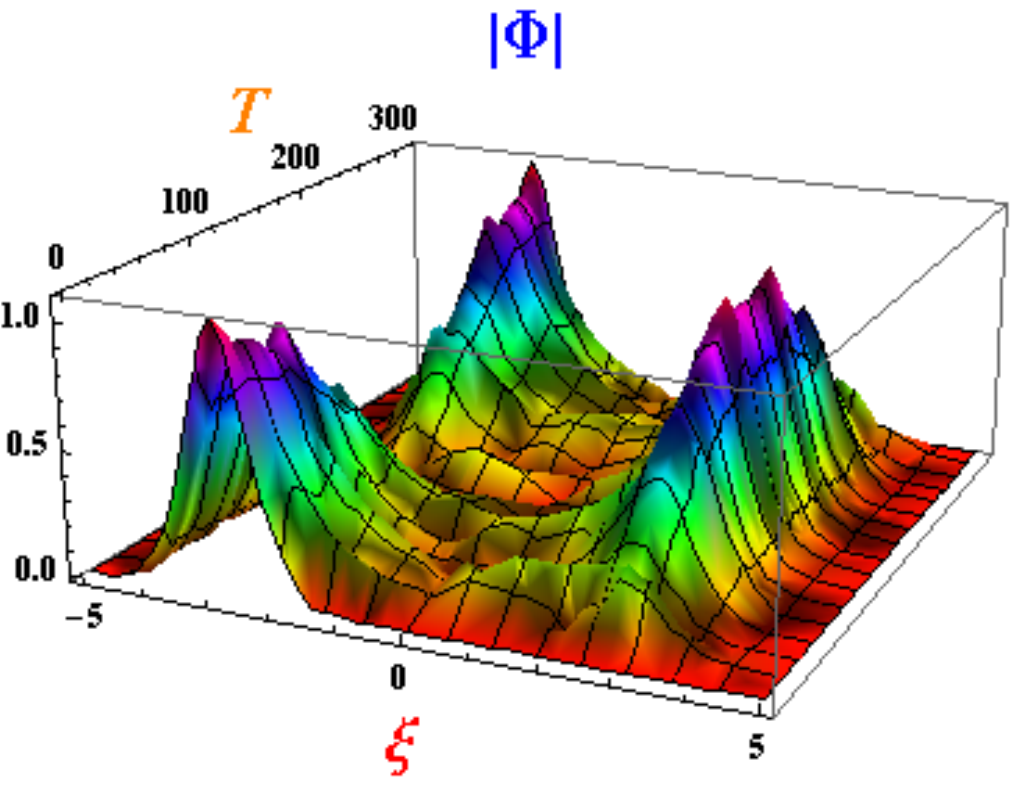} \\ a)}
\end{minipage}
\hfill
\begin{minipage}[h]{0.499\linewidth}
\center{\includegraphics[width=1.0\linewidth]{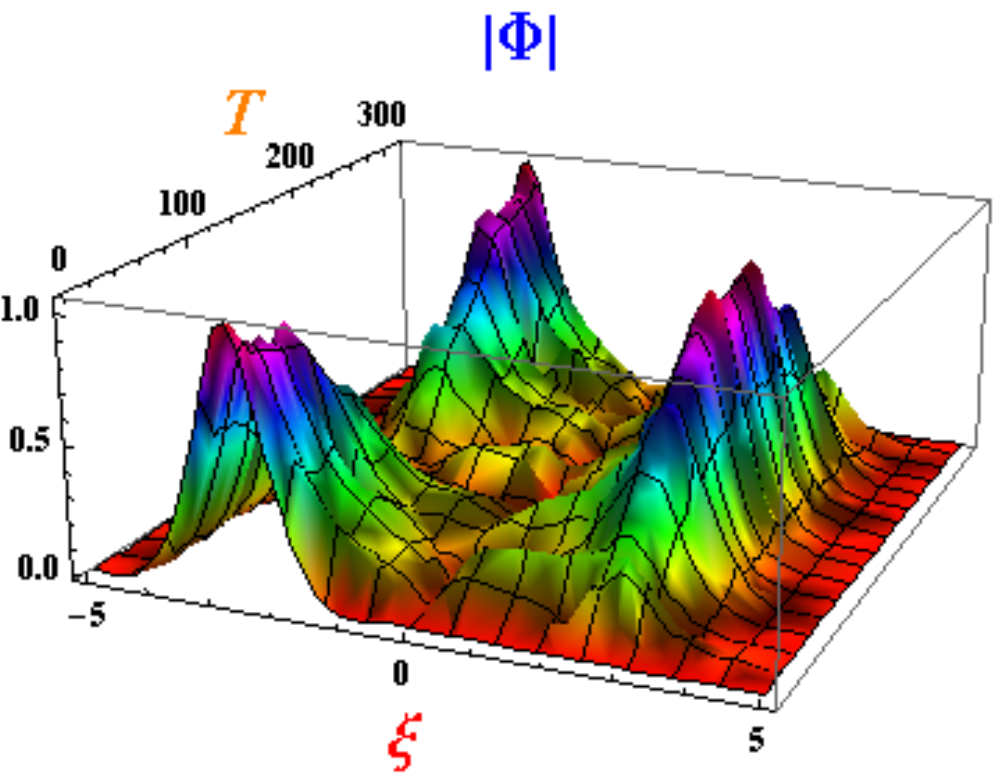} \\ b)}
\end{minipage}
\caption{$|\Phi(\xi,T)|$ dependence in the symmetric case: plot a) corresponds to the third row of table \rf{tab1} and plot b) corresponds to the fourth row of table \rf{tab1}.   }
\label{fig:image5}
\end{figure}
The space-time evolution of the wave packet is characterized by a very small portion of $|\Phi(0,T)|$ in the case and by the independence of the revival time on the choice of the ground-state energy. In other words, a transport of `particles' from side to side wells, with negligibly small central well filling, is realized in this case. The transport time is $1/2\, T_{revival}=\pi/\D$, and the complete returning time of the packet is equal to $T_{revival}$, due to time reversibility of quantum mechanics. To avoid the wave packet returning, laser beam action on a quantum-mechanical system was proposed in \rcite{Creentree04,Eckert04,Opartny09}. Clearly, this mechanism breaks the time invariance. Note also that the central well filling rate decreases with increasing $|\m|$. Hence, changes in $|\m|$ enable us to control the rate of the central well filling in the tunneling processes.

\subsec{Asymmetric case}

Let us consider the effect of deformation of the potential on the tunneling dynamics of wave packets. We suppose $\L_1=1$ and vary the potential shape by changing $\L$. As has been noted above (see figure \rf{fig:image1}), the most significant changes in the shape are achieved at $\L \ra 0$. Table \rf{tab2} contains the coefficients of the series expansion of $\Phi(\xi,0)$ over the states of $\Ht^+_+(\xi,\m,\n,\L_1=1,\L=0.05)$ (recall that here $\w=1$).  

\begin{table}[h] \hspace{-1cm}
			\begin{tabular}{|c|c|c|c|c|c|c|c|c|c|}
		\hline
			state number & 1  & 2 & 3 & 4 & 5 & 6 & 7 & 8 & 9 \\ \hline
			 $\L=0.05,R=0.6$, Left & $0.768$ & $0.21$ & $- 0.564$ & $0.034$ & $-0.026$ & $0.016$ & $-0.006$ & $0.005$ & $-0.022$ \\ 
			  $\m=-0.03,\n=-0.02$ &  &  &  &  &  &  &  &  &  \\ \hline
			$\L=0.05,R=0.8$, Right & $0.17$ & $-0.96$ & $- 0.128$ & $-0.015$ & $-0.014$ & $-0.011$ & $-0.007$ & $0.009$ & $-0.021$ \\ 
			$\m=-0.03,\n=-0.02$ &  &  &  &  &  &  &  &  &  \\ \hline
			$\L=0.05,R=0.3$, Left & $0.078$ & $0.209$ & $- 0.938$ & $0.136$ & $-0.072$ & $0.053$ & $-0.074$ & $0.11$ & $-0.12$ \\ 
			$\m=-1,\n=-0.02$ &  &  &  &  &  &  &  &  &  \\ \hline
			$\L=0.05,R=0.5$, Right & $0.012$ & $-0.95$ & $- 0.215$ & $-0.044$ & $-0.029$ & $-0.012$ & $-0.005$ & $-0.02$ & $0.059$ \\  
			$\m=-1,\n=-0.02$ &  &  &  &  &  &  &  &  &  \\ \hline
		\end{tabular}
		\caption{\label{tab2} Coefficients of the series expansion of $\Phi(\xi,0)$ (cf \rf{Phiapp}) over the states of $\Ht^+_+(\xi,\m,\n,\L_1,\L)$ ($\L_1=1$).}
		\label{tab2}
	\end{table}
Note, that a good approximation of $\Phi(\xi,0)$ is achieved by taking into account the ten lowest states of $\Ht^+_+$. Other states contribute no more than 5\% of probability density of the initial wave packet for different values of parameters, satisfying the potential and the wave packet. Hence, we conclude that the proposed basis is relevant to describe the dynamics of localized states even in the case of a strong deformation of a three-well potential.

The space-time evolution, described in equations \rf{Phidiml}, \rf{Kt++HO}, \rf{K0THO} and \rf{Psit01xi} for a deformed potential, shows a number of differences in comparison to the case of a symmetric potential. First of all, the tunneling transition dynamics is characterized by the partial trapping of the wave packet at the initial well (see figure \rf{fig:image6}). This effect consists in suppressing tunneling transitions in another side well, and it is strongly pronounced when the wave packet is initially situated in the deeper well. The mechanism of the partial `trapping' of the wave packet in the initial well is quite simple. If the maximal contribution to $\Phi(\xi,0)$ located at one of the side wells comes from the under-barrier state, the value of its wave function in the other side well is very small. Hence, the contribution of this state to the tunneling to the other side well process is insignificant. Other under-barrier states give small input into tunneling transitions, due to their minor contributions in forming $\Phi(\xi,0)$.
\begin{figure}[h!]
\begin{minipage}[h]{0.49\linewidth}
\center{\includegraphics[width=0.9\linewidth]{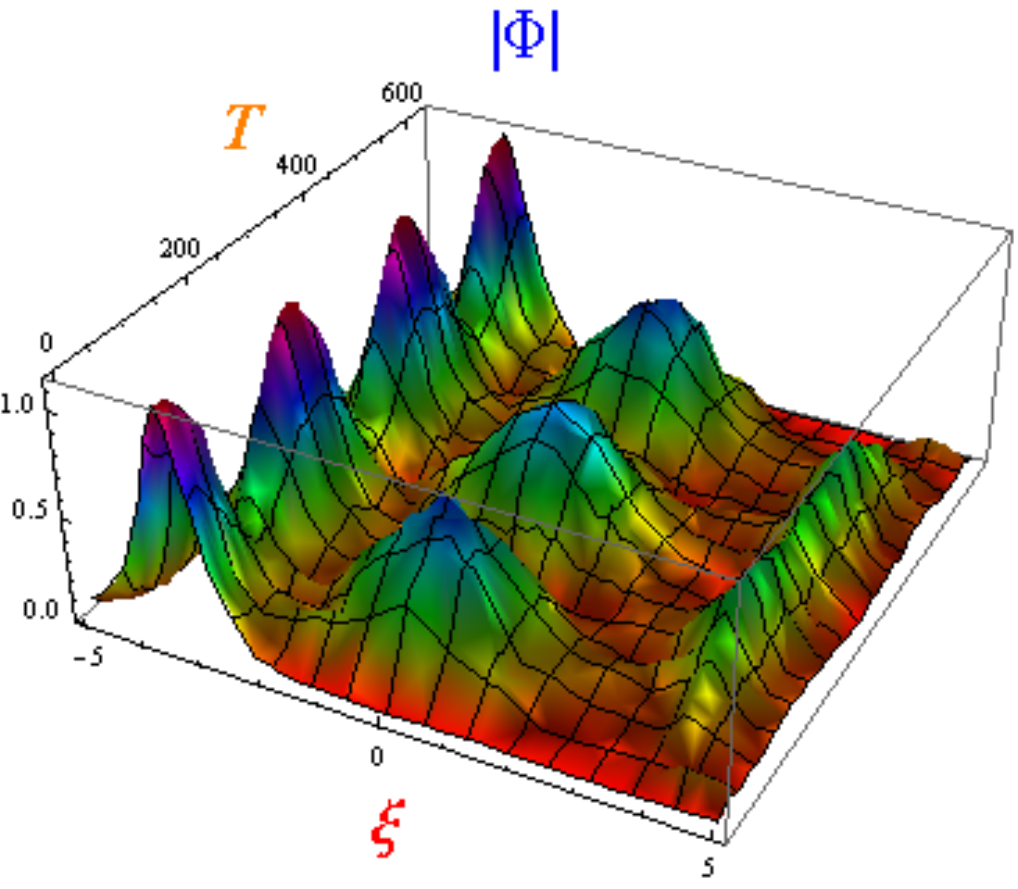} \\ a)}
\end{minipage}
\hfill
\begin{minipage}[h]{0.49\linewidth}
\center{\includegraphics[width=0.99\linewidth]{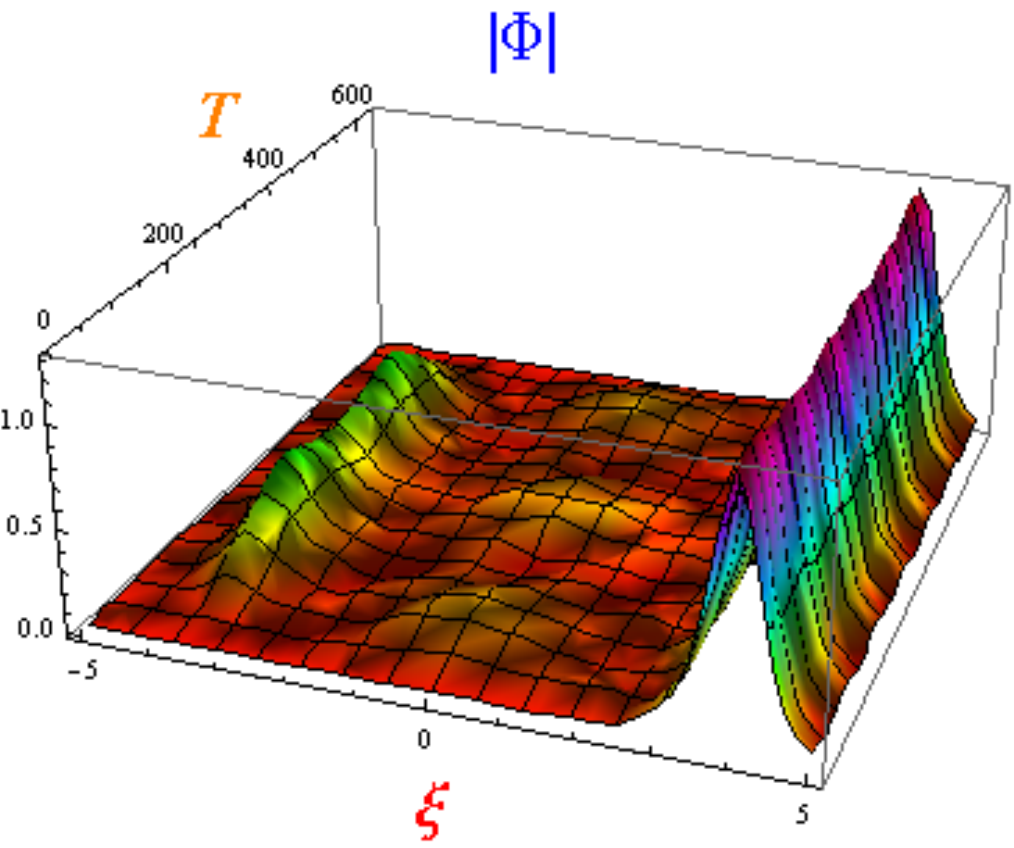} \\ b)}
\end{minipage}
\caption{$|\Phi(\xi,T)|$ dependence in the asymmetric case: plot a) corresponds to the first row of table \rf{tab2} and plot b) corresponds to the second row of table \rf{tab2}.   }
\label{fig:image6}
\end{figure}

Concerning the central well filling, in the case of the initial location of $\Phi(\xi,0)$ in the left well (see the left panel in figure \rf{fig:image6}) its value is big, while in the case of location of $\Phi(\xi,0)$ in the right well, the central well filling value is small (see the right panel in figure \rf{fig:image6}). The same situation is observed for $|\m|\sim 1$ (see Fig.\rf{fig:image7}), though the portion of the wave packet at the central well is negligibly small even when the wave packet is initially located at the left well (compare with the previously considered left panel in figure \rf{fig:image6}).  

\begin{figure}[h!]
\begin{minipage}[h]{0.499\linewidth}
\center{\includegraphics[width=1.1\linewidth]{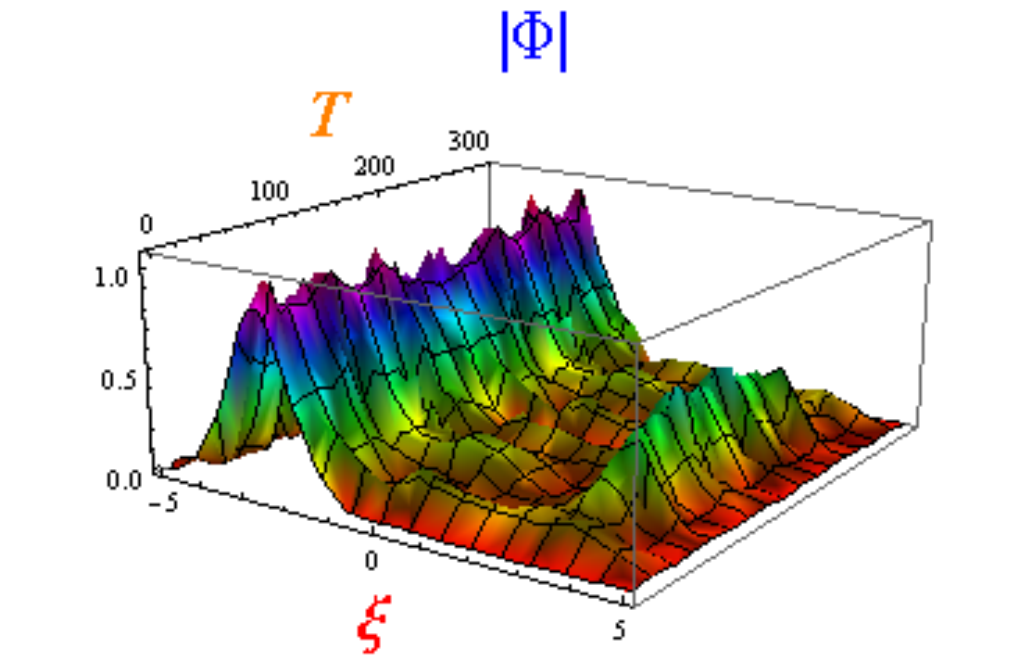} \\ a)}
\end{minipage}
\hfill
\begin{minipage}[h]{0.499\linewidth}
\center{\includegraphics[width=1.1\linewidth]{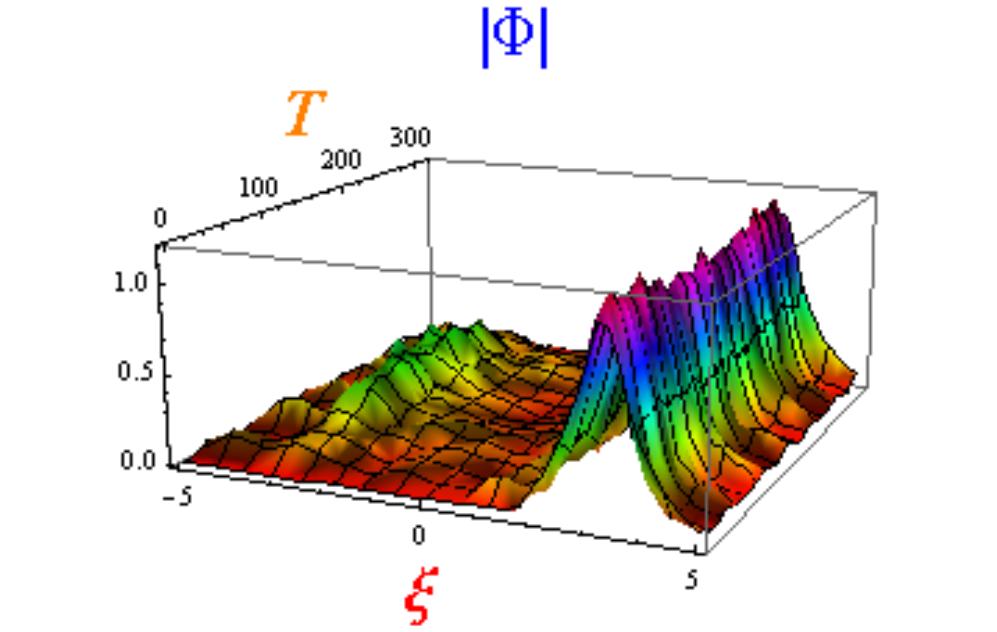} \\ b)}
\end{minipage}
\caption{$|\Phi(\xi,T)|$ dependence in the asymmetric case: Plot a) corresponds to the 3rd row of Table \rf{tab2}; Plot b) corresponds to the 4th row of Table \rf{tab2}.   }
\label{fig:image7}
\end{figure}

As for the case of a symmetric potential, the dynamics of tunneling transitions strongly depends on small changes of the tunnel doublet value $\D=1/2-\ve/\w=-\n$. In particular, it leads to an increase in the period of  the wave packet revival time $T_{revival}=2\pi \w/(E_0-\ve)$.

\newsec{Conclusions}

In this paper, we have developed the previously proposed approach \rcite{Berezovoj12},  based on exactly solvable quantum-mechanical models with multi-well potentials and the corresponding exact propagators, to study the dynamics of initially localized states in triple-well potentials. By the use of the procedure of constructing isospectral N = 4 SQM Hamiltonians with multi- well potentials\footnote{It is noteworthy to mention other recent applications of SQM to condensed matter physics problems \cite{Correa08},\cite{Arancibia12PRD},\cite{Jakubsky11},\cite{Jakubsky12},\cite{Arancibia12}. }, the general expressions of Hamiltonians with three-well potentials and their wavefunctions have been obtained. Parameters satisfying the obtained expressions enable us to change the shapes of three-well potentials in a wide region. We have obtained general expressions for the quantum-mechanical propagators, which take into account all states of the Hamiltonians.

Specifying the initial Hamiltonian to that of the harmonic oscillator model, we applied the results to study features of the tunneling dynamics. It is important to note the relevance of the basis that has been used in the series expansion of the initial wave packet $\Phi(\xi,0)$ to this end. This basis is formed by states of the obtained Hamiltonians under an arbitrary deformation of the potentials. Even in the case of a large squeeze degree of the wave packet, taking the ten lowest states is sufficient to approximate $\Phi(\xi,T)$ in a correct way. This claim has been verified by comparing \rf{Phiapp} to the one computed with the exact propagator \rf{Phidiml}.

The tunneling dynamics in a symmetric three-well potential demonstrates a number of interesting properties. When the tunnel doublet states, which are formed by the first and the second excited levels, are far from the ground state, the wave packet $\Phi(\xi,T)$ demonstrates a `Josephson'-type behavior. The revival time period $T_{revival}$ is determined by the value of the tunnel doublet $\D=(E_0-\ve)/\w$ in the case. Furthermore, the central well filling is small during the tunneling transport from side to side wells, and the filling increases with the growth of $|\ve_1|$. Therefore, by changing the ground-state level, one may control the central well filling during the tunneling transport.

When the three lowest states are close, the central well filling becomes large. The wave packet revival time $T_{revival}$, as well as its qualitative behavior, strongly depends on changes of $\D_1=(\ve-\ve_1)/\w$. The non-regularity in the behavior of $T_{revival}$ is explained by its dependence as on $\D^{-1}$, as well as on $\D_1^{-1}$. The value of $T_{revival}$ is defined by common multiple of $\D^{-1}$ and $\D_1^{-1}$. Hence, small deviations in one of them (say, in $\D_1$) result in the significant increasing $T_{revival}$, which leads to the essential changes in $|\Phi(\xi,T)|$ behavior. Since such a non-monotonic dependence was obtained in the framework of supersymmetric quantum mechanics, it is potentially interesting to figure out this phenomenon in other supersymmetric systems, e.g., in \cite{Arancibia12}.

The tunneling dynamics of wave packets in the deformed three-well potentials demonstrates, as in the case of two-well potentials \rcite{Berezovoj12}, the effect of partial trapping of the wave packet in the original well. Putting it differently, the portion of the wave packet tunneling from one side well to another side well is sufficiently small, independent of the energies of lowest states. The central well filling is essential only when the lowest states are close to each other, and when the wave packet is initially located at a less deep well. Tunneling in the deformed potentials possesses a non-regularity of $|\Phi(\xi,T)|$ too.

Studies of tunneling in atomtronic devices are essentially based on a time-dependent interaction, provided, e.g., by the laser-beam radiation. The effects of having such interactions were investigated in systems with two-well potentials \rcite{Grossman91},\rcite{Lin92},\rcite{Utermann94},\rcite{Kierig08},\rcite{Lizuain09}, where it was demonstrated that the efficiency of the tunneling dynamics is controlled by the use of external driving forces. For systems with three-well potentials, theoretical modeling of the effects of driving forces is at an early stage and is based on matrix Hamiltonian models \rcite{Chen12},\rcite{Lizuain09}, or on potentials, the properties of which can be studied by computer simulations \rcite{Lu10}, \rcite{Lu11}. Handling exactly solvable models with three-well potentials with widely varied shapes, and with the corresponding exact propagators, opens up the possibility to study the tunneling dynamics in time-dependent external forces. We believe that the obtained results will be useful in filling a gap in these investigations.

\ack{We are thankful to Yu.L. Bolotin and V.A. Cherkaskiy for valuable discussions, and to G.I. Ivashkevych for his contribution at early stage of this project. The work of VPB and AJN was supported in part by DFFD-RFFR grant \# F40.2/040. 

}

\end{document}